\documentclass[11pt, prd, aps, showpacs, superscriptaddress, floatfix, nofootinbib]{revtex4-1}
\usepackage{amsmath,amssymb}
\usepackage{geometry}
\usepackage{graphicx}
\usepackage{amssymb}
\usepackage{epstopdf}
\usepackage{bbold}

\newcommand{\be}{\begin{equation}}
\newcommand{\ee}{\end{equation}}
\newcommand{\bea}{\begin{eqnarray}}
\newcommand{\eea}{\end{eqnarray}}

\newcommand{\LaSapienza}{Physics Department and INFN Sezione di Roma La Sapienza,\\ Piazzale Aldo Moro 5, 00185 Roma, Italy}
\newcommand{\RomatreINFN}{Istituto Nazionale di Fisica Nucleare, Sezione di Roma Tre,\\ Via della Vasca Navale 84, I-00146 Rome, Italy}
\newcommand{\SNS}{Scuola Normale Superiore, Piazza dei Cavalieri 7, I-56126, Pisa, Italy}
\newcommand{\INFNpisa}{Istituto Nazionale di Fisica Nucleare, Sezione di Pisa,\\ Largo Bruno Pontecorvo 3, I-56127 Pisa, Italy}

\begin{document}

\title{Constraints for the semileptonic $B \to D^{(*)}$ form factors\\[2mm] from lattice QCD simulations of two-point correlation functions}

\author{G.~Martinelli}\affiliation{\LaSapienza}
\author{S.~Simula}\affiliation{\RomatreINFN}
\author{L.~Vittorio}\affiliation{\SNS},\affiliation{\INFNpisa}

%\pacs{11.15.Ha, % Lattice gauge theory
%         12.15.Ff, % Quark and lepton masses and mixing
%         12.38.Gc, % Lattice QCD calculations
%         13.20.-v	, %Leptonic, semileptonic, and radiative decays of mesons
%}

\begin{abstract}
In this work we present the first non-perturbative determination of the hadronic susceptibilities that constrain the form factors entering the semileptonic $B \to D^{(*)} \ell \nu_\ell $ transitions due to unitarity and analyticity. The susceptibilities are obtained by evaluating moments of suitable two-point correlation functions obtained on the lattice. Making use of the gauge ensembles produced by the Extended Twisted Mass Collaboration with $N_f = 2+1+1$ dynamical quarks at three values of the lattice spacing ($a \simeq 0.062, 0.082, 0.089$ fm) and with pion masses in the range $\simeq 210 - 450$ MeV, we evaluate the longitudinal and transverse susceptibilities of the vector and axial-vector polarization functions at the physical pion point and in the continuum and infinite volume limits. The ETMC ratio method is adopted to reach the physical $b$-quark mass $m_b^{phys}$. At zero momentum transfer for the $b \to c$ transition we get $\chi_{0^+}(m_b^{phys}) = 7.58\,(59) \cdot 10^{-3}$, $\chi_{1^-}(m_b^{phys}) = 6.72\,(41) \cdot 10^{-4}$ GeV$^{-2}$, $\chi_{0^-}(m_b^{phys}) = 2.58\,(17) \cdot 10^{-2}$ and $\chi_{1^+}(m_b^{phys}) = 4.69\,(30) \cdot 10^{-4}$ GeV$^{-2}$ for the scalar, vector, pseudoscalar and axial susceptibilities, respectively. In the case of the vector and pseudoscalar channels the one-particle contributions due to $B_c^*$- and $B_c$-mesons are evaluated and subtracted to improve the bounds, obtaining: $\chi_{1^-}^{sub}(m_b^{phys}) = 5.84\,(44) \cdot 10^{-4}$ GeV$^{-2}$ and $\chi_{0^-}^{sub}(m_b^{phys}) = 2.19\,(19) \cdot 10^{-2}$.

\end{abstract}

\maketitle

\section{Introduction}
\label{sec:intro}

A primary goal of flavor physics is the precise determination of the elements of the Cabibbo-Kobayashi-Maskawa (CKM) mixing matrix~\cite{CKM}, since its unitarity represents an important probe for the possible presence of physics beyond the Standard Model (SM). 
Consequently, efforts have been devoted to both the experimental and the theoretical investigations of inclusive and exclusive semileptonic decays of hadrons.

In the case of the CKM entry $V_{cb}$, describing the weak mixing between the bottom and the charm quarks, there is a tension between the two determinations coming from inclusive and exclusive semileptonic $B$-meson decays~\cite{HFLAV}.
Till now such a tension has not received a satisfactory explanation~\cite{FLAG,PDG,Gambino:2020jvv,Crivellin:2014zpa}.
Moreover, measurements of the exclusive branching ratio $B \to D^{(*)} \tau \nu$ over  $B \to D^{(*)} \ell \nu_\ell$ with $\ell = e, \mu$ made at Belle, Babar, and LHCb~\cite{Lees:2012xj,Lees:2013uzd,Aaij:2015yra,Huschle:2015rga,Sato:2016svk,Hirose:2016wfn,Aaij:2017uff,Hirose:2017dxl,Aaij:2017deq}  show discrepancies with the SM predictions, suggesting a possible violation of lepton flavor universality~\cite{Bernlochner:2017jka,Bernlochner:2017xyx,Jung:2018lfu,Colangelo:2018cnj,Azatov:2018knx,Feruglio:2018fxo}. 

In the last few years the exclusive semileptonic $B \to D^{(*)} \ell \nu_\ell$ decays have received significant attention.
As well known, the crucial ingredients for the extraction of $|V_{cb}|$ are the hadronic form factors entering the exclusive $B$-meson decays. 
Reliable information on the latter is given by first-principle calculations performed using QCD simulations on the lattice.
However, since the bottom quark is so heavy, its mass in lattice units is still larger than one on the currently available lattice setups.
Thus, cutoff effects (as well as large statistical fluctuations) are a limiting factor, which makes very difficult to determine on the lattice the dependence of the form factors on the squared 4-momentum transfer $q^2$, where $q = p_B - p_{D^{(*)}}$ is the lepton-pair 4-momentum, in the full kinematical range $m_\ell^2 \leq q^2 \leq (m_B - m_{D^{(*)}})^2$.
As a matter of fact, the results for the $B \to D \ell \nu_\ell$ form factors from lattice QCD~\cite{Na:2015kha,Lattice:2015rga} are available only in a limited range of values of $q^2$, namely $9.3~{\rm GeV}^2 \lesssim q^2 \lesssim 11.7~{\rm GeV}^2$, much smaller than the full physical range, $0 \lesssim q^2 \lesssim 11.7~{\rm GeV}^2$. 
Also preliminary results for the momentum dependence of the form factors entering the $B \to D^* \ell \nu$ decays are still restricted at small recoil, i.e.~at large values of $q^2$~\cite{Kaneko:2019vkx,Aviles-Casco:2019zop}. 

In order to supply the lack of information at large recoil, i.e.~at small values of $q^2$, both experimental and theoretical analyses have to parameterise the form factors in order to describe their momentum dependence in the full kinematical range.
In this way, the extraction of $|V_{cb}|$ from experiments may be biased by the theoretical model adopted for fitting the data. 
In the years most of the analyses used two popular parameterisations, called Boyd-Grinstein-Lebed (BGL)~\cite{ Boyd:1995cf,Boyd:1995sq,Boyd:1997kz} or Caprini-Lellouch-Neubert (CLN)~\cite{Caprini:1995wq,Caprini:1997mu}.
Recent experimental results~\cite{Waheed:2018djm,Dey:2019bgc} for the momentum and angular distributions have allowed studies of the role played by the BGL and CLN parameterisations of the semileptonic form factors~\cite{Bigi:2016mdz,Grinstein:2017nlq,Bigi:2017njr,Bigi:2017jbd,Gambino:2019sif,Iguro:2020cpg}.
It turned out that the sensitivity to the specific parametrization employed can be solved only by the precise knowledge of the form factors at non-zero recoil. 

A possible way to constrain in a model-independent way the $q^2$-dependence of the form factors extracted from lattice QCD along the full kinematical range  has been proposed recently in Ref.~\cite{DiCarlo:2021dzg}.
Starting from the pioneering work of Ref.~\cite{Lellouch:1995yv} the dispersive matrix (DM) method of Ref.~\cite{DiCarlo:2021dzg} makes use of unitarity and analyticity bounds~\cite{Boyd:1995cf,Boyd:1995sq,Boyd:1997kz} applied to lattice data and introduces a formalism to take into account the uncertainties of the lattice results. 
The DM method contains several new elements with respect to the original proposal of Ref.~\cite{Lellouch:1995yv} and allows to extract, using suitable two-point functions computed non-perturbatively, the form factors at low values of $q^2$ from those computed explicitly on the lattice at large $q^2$, without any assumption about their $q^2$-dependence.
The DM method was tested in the case of the semileptonic $D \to K \ell \nu_\ell$ decay, by comparing its results with the explicit direct lattice calculation of the form factors available in the full $q^2$-range. 
It was shown~\cite{DiCarlo:2021dzg} that the DM method is very effective and allows to compute the form factors in a model-independent way and with rather good precision in the low-$q^2$ region not accessible directly to lattice calculations. 

Our aim is to apply the DM method of Ref.~\cite{DiCarlo:2021dzg} to the case of exclusive semileptonic $B$-meson decays.
The DM method requires the non-perturbative evaluation of moments of suitable two-point correlation functions and the subtraction of non-perturbative single-particle contributions. These are essential ingredients of the whole strategy and they represent per se a complex step, since it involves the determination of  single-particle contributions to the two-point functions and, for heavy quarks, the control of large discretization effects.
In this work we address the determination of the subtracted susceptibilities in the case of the $b \to c$ transitions, while the application of the DM method to the extraction of $|V_{cb}|$ from the experimental data on the exclusive $B \to D^{(*)} \ell \nu_\ell$ decays will be the subject of a separate work~\cite{Martinelli:2021onb}.

The two-point correlation functions are evaluated making use of the gauge ensembles produced by the Extended Twisted Mass Collaboration (ETMC) with $N_f = 2+1+1$ dynamical quarks at three values of the lattice spacing ($a \simeq 0.062, 0.082, 0.089$ fm) and with pion masses in the range $\simeq 210 - 450$ MeV~\cite{Baron:2010bv,Baron:2011sf}. Our results are extrapolated to the continuum limit and to the physical pion point. In order to reduce cutoff effects we take advantage of the calculation of the two-point correlation functions carried out in perturbation theory both in the continuum and on the lattice in Ref.~\cite{DiCarlo:2021dzg}.
For reaching the $b$-quark point we employ the ETMC ratio method of Ref.~\cite{Blossier:2009hg}, that was already applied to determine the  physical $b$-quark mass, $m_b^{phys}$, on the same gauge ensembles in Ref.~\cite{Bussone:2016iua}. 

In terms of the scalar ($0^+$), vector ($1^-$), pseudoscalar ($0^-$) and axial ($1^+$) susceptibilities, which at zero momentum transfer are moments of the appropriate two-point correlation functions, we get the following results for the $b \to c$ transition
 \bea
     \label{eq:chi0+_intro}
     \chi_{0^+}(m_b^{phys}) & = & 7.58\,(59) \cdot 10^{-3} ~ , ~ \\[2mm]
     \label{eq:chi1-_intro}
     \chi_{1^-}(m_b^{phys}) & = & 6.72\,(41) \cdot 10^{-4}~{\rm GeV}^{-2} ~ , ~ \\[2mm]
     \label{eq:chi0-_intro}
     \chi_{0^-}(m_b^{phys}) & = & 2.58\,(17) \cdot 10^{-2} ~ , ~ \\[2mm]
     \label{eq:chi1+_intro}
     \chi_{1^+}(m_b^{phys}) & = & 4.69\,(30) \cdot 10^{-4}~{\rm GeV}^{-2} ~ . ~
 \eea
In Refs.~\cite{Bigi:2016mdz,Bigi:2017njr,Bigi:2017jbd} the susceptibilities have been estimated using perturbation theory (PT) at next-to-next-to-leading order (NNLO), obtaining: $\chi_{0^+}(m_b^{phys}) = 6.204~(81) \cdot 10^{-3}$,  $\chi_{1^-}(m_b^{phys}) = 6.486~(48) \cdot 10^{-4}$ GeV$^{-2}$, $\chi_{0^-}(m_b^{phys}) = 2.41 \cdot 10^{-2}$ and $\chi_{1^+}(m_b^{phys}) = 3.894 \cdot 10^{-4}$ GeV$^{-2}$.
While the vector and pseudoscalar susceptibilities are only $\simeq 4 \%$ and $\simeq 7 \%$ lower than our findings (\ref{eq:chi1-_intro}) and (\ref{eq:chi0-_intro}), the scalar and axial susceptibilities are $\simeq 20 \%$ lower than our results (\ref{eq:chi0+_intro}) and (\ref{eq:chi1+_intro}), respectively. 
In all cases the differences are within $\sim 2.5$ standard deviations.

In order to improve the bounds of the DM method, in the case of the vector and pseudoscalar channels the one-particle contributions due to $B_c^*$- and $B_c$-mesons are evaluated and subtracted from Eqs.~(\ref{eq:chi1-_intro}) and (\ref{eq:chi0-_intro}), respectively, obtaining
\bea
    \label{eq:chi1-_sub_intro}
     \chi_{1^-}^{sub}(m_b^{phys}) & = & 5.84\,(44) \cdot 10^{-4}~{\rm GeV}^{-2} ~ , ~ \\[2mm]
    \label{eq:chi0-_sub_intro}
     \chi_{0^-}^{sub}(m_b^{phys}) & = & 2.19\,(19) \cdot 10^{-2} ~ . ~
\eea

Eqs.~(\ref{eq:chi0+_intro}), (\ref{eq:chi1+_intro}), (\ref{eq:chi1-_sub_intro}) and (\ref{eq:chi0-_sub_intro}) represent our non-perturbative determinations of the  scalar, axial-vector, vector and pseudoscalar susceptibilities entering the dispersive bounds on the form factors of the exclusive semileptonic $B \to D^{(*)} \ell \nu_\ell$ decays. 

The plan of the paper is as follows.
In Section~\ref{sec:euclidean} we give the basic definitions of the the quantities relevant in this work, namely the longitudinal and transverse susceptibilities, which are moments of suitable two-point correlation functions.
In Section~\ref{sec:susceptibilities} we describe the lattice calculations of the latter ones using the gauge ensembles produced by the ETMC with $N_f = 2+1+1$ dynamical quarks (see Appendix~\ref{sec:simulations}) for the $h \to c$ transition, where $h$ is a quark heavier than the charm. We illustrate the presence of contact terms in the longitudinal susceptibilities and how to avoid them by means of Ward Identities (WIs). We apply the perturbative calculations of Ref.~\cite{DiCarlo:2021dzg} to understand the main features of the contact terms and also to get a beneficial reduction of the cutoff effects on the susceptibilities.
In Section~\ref{sec:ETMC_ratio} we apply the ETMC ratio method of Ref.~\cite{Blossier:2009hg} to reach the physical $b$-quark point. We first construct suitable ratios of the susceptibilities, evaluated at two subsequent values of the heavy-quark mass $m_h$, and perform the extrapolation to the continuum limit and to the physical pion point. Then, the resulting ratios are smoothly interpolated at the physical $b$-quark mass, determined in Ref.~\cite{Bussone:2016iua}, taking advantage of the fact that the ratios are defined in such a way to guarantee that the their values are exactly known in the heavy-quark limit. 
In Section~\ref{sec:grs} we evaluate the contributions of the $B_c^*$- and $B_c$-mesons to the vector and pseudoscalar susceptibilities, respectively, obtaining in this way a more stringent bound for the DM method of Ref.~\cite{DiCarlo:2021dzg}.
Finally, Section~\ref{sec:conclusions} is devoted to our conclusions and outlooks for future developments.

\section{Euclidean two-point correlation functions}
\label{sec:euclidean}

In this Section we briefly recall the definition of the longitudinal and transverse susceptibilities, which are the quantities relevant in this work.
More details can be found in Ref.~\cite{DiCarlo:2021dzg}. 

Let's start with the vacuum polarization tensors $\Pi_{\mu \nu}^{V(A)}$ related to the product of vector and axial-vector currents of the the flavor changing $b \to c$ weak transition~\cite{Boyd:1997kz,Caprini:1997mu}. 
Performing a Wick rotation from Minkowskian coordinates to Euclidean ones, $x \equiv (t, \vec{x})$, the vacuum polarization tensors can be written as~\cite{DiCarlo:2021dzg}
\bea
      \label{eq:vector}
      \Pi_{\mu \nu}^V(Q) & = & \int d^4x ~ e^{-iQ \cdot x} \langle 0 | T\left[ \bar{b}(x) \gamma_\mu^E c(x) ~ 
                                               \bar{c}(0) \gamma_\nu^E b(0) \right] | 0 \rangle \nonumber \\ 
                                     & = &  - Q_\mu Q_\nu \Pi_{0^+}(Q^2) + (\delta_{\mu \nu} Q^2 - Q_\mu Q_\nu) \Pi_{1^-}(Q^2) ~ , \\[2mm]
      \label{eq:axial}
      \Pi_{\mu \nu}^A(Q) & = & \int d^4x ~ e^{-iQ \cdot x} \langle 0 | T\left[ \bar{b}(x) \gamma_\mu^E \gamma_5^E c(x) ~ 
                                               \bar{c}(0) \gamma_\nu^E \gamma_5^E b(0) \right] | 0 \rangle \nonumber \\ 
                                     & = &  - Q_\mu Q_\nu \Pi_{0^-}(Q^2) + (\delta_{\mu \nu} Q^2 - Q_\mu Q_\nu) \Pi_{1^+}(Q^2) ~ ,
\eea 
where $\gamma_\mu^E$ are (Hermitean) Euclidean Dirac matrices (i.e., $\gamma_0^E = \gamma_0^M$, $\vec{\gamma}^E = -i \vec{\gamma^M}$ and $\gamma_5^E = \gamma_5^M$) and $Q \equiv (Q_0, \vec{Q})$  is an Euclidean 4-momentum (i.e., $Q^2 \equiv Q_0^2 + |\vec{Q}|^2$).
In Eqs.~(\ref{eq:vector})-(\ref{eq:axial}) the quantities $\Pi_{0^{\pm}}$ and $\Pi_{1^{\mp}}$ are called the polarization functions and the subscirpt identifies the spin-parity of the various channels.
In particular, the term proportional to $\Pi_{0^{+}}$ ($\Pi_{0^{-}}$) represents the \emph{longitudinal} part of the polarization tensor with vector (axial) four-currents, while the term proportional to $\Pi_{1^{-}}$ ($\Pi_{1^{+}}$) is the \emph{transverse} contribution to the polarization tensor with vector (axial) four-currents.
In what follows, for sake of simplicity, we will omit the explicit superscript $E$ in the definition of the Euclidean $\gamma$-matrices.
 
Choosing the four-momentum $Q$ in the temporal direction $Q = (Q, \vec{0})$ one has
 \be
       \label{eq:VLVTALAT}
       Q^2 ~ \Pi_{(0^+, 1^-, 0^-, 1^+)}(Q^2) = - \int_{-\infty}^\infty dt ~ e^{-iQt} ~ C_{(0^+, 1^-, 0^-, 1^+)}(t) ~ , ~
\ee
where the Euclidean correlators $C_{(0^+, 1^-, 0^-, 1^+)}(t)$ are given by
\bea
    \label{eq:CVL}
    C_{0^+}(t) & = & \int d^3x  \langle 0 | T\left[ \bar{b}(x) \gamma_0 c(x) ~ \bar{c}(0) \gamma_0 b(0) \right] | 0 \rangle ~ , \\[2mm]
    \label{eq:CVT}
    C_{1^-}(t) & = & \frac{1}{3} \sum_{j=1}^3 \int d^3x  \langle 0 | T\left[ \bar{b}(x) \gamma_j c(x) ~ \bar{c}(0) \gamma_j b(0) \right] | 0 \rangle ~ , \\[2mm]
   \label{eq:CAL}
    C_{0^-}(t) & = & \int d^3x  \langle 0 | T\left[ \bar{b}(x) \gamma_0 \gamma_5 c(x) ~ \bar{c}(0) \gamma_0 \gamma_5 b(0) \right] | 0 \rangle ~ , \\[2mm]
    \label{eq:CAT}
    C_{1^+}(t) & = & \frac{1}{3} \sum_{j=1}^3 \int d^3x \langle 0 | T\left[ \bar{b}(x) \gamma_j \gamma_5 c(x) ~ 
                               \bar{c}(0) \gamma_j \gamma_5 b(0) \right] | 0 \rangle ~ .
\eea
The quantities relevant in this work are the susceptibilities $\chi(Q^2)$, which are either first or second derivatives of the polarization functions $\Pi(Q^2)$, namely
\bea
     \label{eq:chiVL}
       \chi_{0^+}(Q^2) & \equiv & \frac{\partial}{\partial Q^2} \left[ Q^2 \Pi_{0^+}(Q^2) \right] = \int_0^\infty dt ~ t^2 j_0(Qt) ~ C_{0^+}(t) ~ , \\[2mm]
       \label{eq:chiVT}
        \chi_{1^-}(Q^2) & \equiv & - \frac{1}{2} \frac{\partial^2}{\partial^2 Q^2} \left[ Q^2 \Pi_{1^-}(Q^2) \right] = 
                                                   \frac{1}{4} \int_0^\infty dt ~ t^4 \frac{j_1(Qt)}{Qt} ~ C_{1^-}(t) ~ , \\[2mm]
       \label{eq:chiAL}
       \chi_{0^-}(Q^2) & \equiv & \frac{\partial}{\partial Q^2} \left[ Q^2 \Pi_{0^-}(Q^2) \right] = \int_0^\infty dt ~ t^2 j_0(Qt) ~ C_{0^-}(t)~ , \\[2mm]
       \label{eq:chiAT}
       \chi_{1^+}(Q^2) & \equiv & - \frac{1}{2} \frac{\partial^2}{\partial^2 Q^2} \left[ Q^2 \Pi_{1^+}(Q^2) \right] = 
                                                   \frac{1}{4} \int_0^\infty dt ~ t^4 \frac{j_1(Qt)}{Qt} ~ C_{1^+}(t) ~ ,
\eea
where $j_0(x) = \mbox{sin}(x) / x$ and $j_1(x) = [\mbox{sin}(x) / x -  \mbox{cos}(x)] / x$ are spherical Bessel functions.
Note that the longitudinal derivatives (\ref{eq:chiVL}) and (\ref{eq:chiAL}) are dimensionless, while the transverse ones (\ref{eq:chiVT}) and (\ref{eq:chiAT}) have the dimension of $[E]^{-2}$, where $E$ is an energy.

Eqs.~(\ref{eq:chiVL})-(\ref{eq:chiAT}) have been obtained in the Euclidean region $Q^2 \geq 0$, but, as shown in Ref.~\cite{DiCarlo:2021dzg}, they can be easily generalized also to the case $Q^2 < 0$.
In the Euclidean region $Q^2 \geq 0$ a good convergence of the perturbative calculation of the above derivatives is expected to occur far from the kinematical regions where resonances can contribute.
In the case of the $b \to c$ weak transition this means down to $Q^2 = 0$~\cite{Boyd:1997kz,Caprini:1997mu}.
Thus, the value $Q^2 = 0$ has been generally employed in the evaluation of the dispersive bounds on heavy-to-heavy \cite{Boyd:1997kz,Caprini:1997mu,Bigi:2016mdz,Bigi:2017njr,Bigi:2017jbd} and heavy-to-light \cite{Lellouch:1995yv,Bourrely:2008za} semileptonic form factors. 
On the contrary, with a non-perturbative determination of the two-point correlation functions we can use the most convenient value of $Q^2$ at disposal, namely the value which will allow the most stringent bounds on the semileptonic form factors~\cite{DiCarlo:2021dzg}. 
In this work we will limit ourselves to the usual choice $Q^2 = 0$, which will allow the comparison with the perturbative results at NNLO obtained in Refs.~\cite{Bigi:2016mdz,Bigi:2017njr,Bigi:2017jbd}, and we will leave the investigation of the choice $Q^2 \neq 0$ to a future work.

At $Q^2 = 0$ the derivatives of the longitudinal and transverse polarization functions correspond to the second and fourth moments of the  longitudinal and transverse Euclidean correlators, $i.e.$
 \bea
     \label{eq:chiVL0}
       \chi_{0^+}(Q^2=0) & = & \int_0^\infty dt ~ t^2 ~ C_{0^+}(t) ~ , \\[2mm]
       \label{eq:chiVT0}
        \chi_{1^-}(Q^2=0) & = & \frac{1}{12} \int_0^\infty dt ~ t^4 ~ C_{1^-}(t) ~ , \\[2mm]
       \label{eq:chiAL0}
       \chi_{0^-}(Q^2=0) & = & \int_0^\infty dt ~ t^2 ~ C_{0^-}(t)~ , \\[2mm]
       \label{eq:chiAT0}
       \chi_{1^+}(Q^2=0) & = & \frac{1}{12} \int_0^\infty dt ~ t^4 ~ C_{1^+}(t) ~ .
\eea

In Ref.~\cite{DiCarlo:2021dzg} it has been shown that the Ward Identities (WIs), which should be satisfied by the vector and axial-vector quark currents, allow to express the longitudinal susceptibilities~(\ref{eq:chiVL0}) and~(\ref{eq:chiAL0}) as the fourth moments of the scalar and pseudoscalar correlation functions, namely one has
\bea
     \label{eq:chiVL_WI0}
     \chi_{0^+}(Q^2 = 0) & = &\frac{1}{12} (m_b - m_c)^2 \int_0^\infty dt ~ t^4 ~ C_S(t) ~ , \quad \\[2mm]
     \label{eq:chiAL_WI0}
     \chi_{0^-}(Q^2 = 0) & = & \frac{1}{12} (m_b + m_c)^2 \int_0^\infty dt ~ t^4 ~ C_P(t) ~ , \quad
\eea
where
\bea
   \label{eq:CS}
    C_S(t) & = & \int d^3x  \langle 0 | T\left[ \bar{b}(x) c(x) ~ \bar{c}(0) b(0) \right] | 0 \rangle ~ , \\[2mm]
    \label{eq:CP}
    C_P(t) & = & \int d^3x  \langle 0 | T\left[ \bar{b}(x) \gamma_5 c(x) ~ \bar{c}(0) \gamma_5b(0) \right] | 0 \rangle ~ .
\eea

\section{Longitudinal and transverse susceptibilities}
\label{sec:susceptibilities}

The gauge ensembles used in this work have been generated by ETMC with $N_f = 2 + 1 + 1$ dynamical quarks, which include in the sea, besides two light mass-degenerate quarks ($m_u = m_d = m_{ud}$), also the strange and the charm quarks with masses close to their physical values~\cite{Baron:2010bv,Baron:2011sf}.
The ensembles are the same adopted to determine the up, down, strange and charm quark masses in Ref.~\cite{Carrasco:2014cwa} and the bottom quark mass in Ref.~\cite{Bussone:2016iua}.
Details are given in Appendix~\ref{sec:simulations}.

Using the ETMC gauge ensembles of Table~\ref{tab:simudetails} we have evaluated the following two-point correlation functions 
\bea
    \label{eq:CVL12}
    C_{0^+}(t) & = & \widetilde{Z}_V^2 ~ \int d^3x  \langle 0 | T\left[ \bar{q}_1(x) \gamma_0 q_2(x) ~ \bar{q}_2(0) \gamma_0 q_1(0) \right] | 0 \rangle ~ , \\[2mm]
    \label{eq:CVT12}
    C_{1^-}(t) & = & \widetilde{Z}_V^2 ~ \frac{1}{3} \sum_{j=1}^3 \int d^3x  \langle 0 | T\left[ \bar{q}_1(x) \gamma_j q_2(x) ~ \bar{q}_2(0) \gamma_j q_1(0) \right] | 0 \rangle ~ , \\[2mm]
   \label{eq:CAL12}
    C_{0^-}(t) & = & \widetilde{Z}_A^2 ~ \int d^3x  \langle 0 | T\left[ \bar{q}_1(x) \gamma_0 \gamma_5 q_2(x) ~ \bar{q}_2(0) \gamma_0 \gamma_5 q_1(0) \right] | 0 \rangle ~ , \\[2mm]
    \label{eq:CAT12}
    C_{1^+}(t) & = & \widetilde{Z}_A^2 ~ \frac{1}{3} \sum_{j=1}^3 \int d^3x \langle 0 | T\left[ \bar{q}_1(x) \gamma_j \gamma_5 q_2(x) ~ \bar{q}_2(0) \gamma_j \gamma_5 q_1(0) \right] | 0 \rangle ~ , \\[2mm]
   \label{eq:CS12}
   C_S(t) & = & \widetilde{Z}_S^2 ~ \int d^3x \langle 0 | T\left[ \bar{q}_1(x) q_2(x) ~  \bar{q}_2(0) q_1(0) \right] | 0 \rangle ~ , \\[2mm]
    \label{eq:CP12}
    C_P(t) & = & \widetilde{Z}_P^2 ~ \int d^3x  \langle 0 | T\left[ \bar{q}_1(x) \gamma_5 q_2(x) ~ \bar{q}_2(0) \gamma_5 q_1(0) \right] | 0 \rangle ~ , ~
\eea
where $q_1$ and $q_2$ are the two valence quarks with bare masses $a \mu_1$ and $a \mu_2$ given in Table~\ref{tab:simudetails}, while the multiplicative factor $\widetilde{Z}_\gamma$ ($\gamma = \{ V, A, S, P \}$) is an appropriate renormalization constant (RC), which will be specified in a while.
Indeed, we consider either opposite or equal values for the Wilson parameters $r_1$ and $r_2$ of the two valence quarks, namely either the case $r_1 = - r_2$ or the case $r_1 = r_2$. 
Since our twisted-mass setup is at its maximal twist, in the case $r_1 = - r_2$ we have $\widetilde{Z}_\gamma = \{ Z_A, Z_V, Z_P, Z_S \}$, while in the case $r_1 = r_2$ we have $\widetilde{Z}_\gamma = \{ Z_V, Z_A, Z_S, Z_P \}$, where the RCs of the various bilinears have been determined in the RI$^\prime$-MOM scheme in Ref.~\cite{Carrasco:2014cwa}.
Once renormalized the correlation functions (\ref{eq:CVL12}-\ref{eq:CP12}) corresponding to either opposite or equal values of the Wilson parameters $r_1$ and $r_2$ differ only by effects of order ${\cal{O}}(a^2)$.

We start by considering the longitudinal and transverse susceptibilities of both the vector and the axial-vector currents $\chi_j(Q^2=0)$ with $j = \{ 0^+, 1^-, 0^-, 1^+ \}$, defined in Eqs.~(\ref{eq:chiVL0}-\ref{eq:chiAT0}) as either the second or the fourth moments of the corresponding longitudinal and transverse Euclidean correlators $C_j(t)$.
For sake of simplicity, in what follows we will indicate by $\chi_j$ the susceptibilities evaluated at $Q^2 = 0$.

For each gauge ensemble the values of $\chi_j$ are obtained for many combinations of the two valence quark masses $m_1 = a \mu_1 / (Z_P a)$ and $m_2 = a \mu_2 / (Z_P a)$, namely for all the 14 values in the light, strange, charm and heavier-than-charm sectors (see Table~\ref{tab:simudetails}) in the case of $a \mu_1$, while the values of $a \mu_2$ have been chosen in the light, strange and charm regions (a total of 7 values).

\subsection{The $h \to c$ transition}
\label{sec:htoc}

In this work we limit ourselves to the quark mass combinations $a \mu_1 = a \mu_h \geq a \mu_c$ and $a \mu_2 = a \mu_c$, which correspond to $h \to c$ transitions.

The values of the simulated susceptibilities $\chi_{0^\pm(1^\pm)}$ are smoothly interpolated at $m_c = a \mu_c / (Z_P a) = m_c^{phys}$ and at a series of values of the heavy-quark mass $m_h = a \mu_h / (Z_P a)$, dictated by the ETMC ratio method (see Ref.~\cite{Bussone:2016iua}), given by
\be
     \label{eq:mh_n}
     m_h(n) = \lambda^{n-1} ~ m_c^{phys} \qquad \mbox{for}~n = 1, 2, ...
\ee
with $\lambda \equiv [m_b^{phys} / m_c^{phys}]^{1/10} = [5.198 / 1.176]^{1/10} \simeq 1.1602$, and starting from $m_h(1) = m_c^{phys}$.
The value of $\lambda$, which is the same as the one adopted in Ref.~\cite{Bussone:2016iua}, is such that $m_h(n = 11) = m_b^{phys}$.
Correspondingly, the uncertainty $\delta m_h(n)$ is given by
\be
     \delta m_h(n) = \epsilon^{n-1} \delta m_c^{phys} \qquad \mbox{for}~n = 1, 2, ...
\ee
with $\epsilon \equiv [\delta m_b^{phys} / \delta m_c^{phys}]^{1/10} = [0.122 / 0.039]^{1/10} \simeq 1.1208$. 
Given the number of simulated values of $m_h > m_c^{phys}$ (see Table~\ref{tab:simudetails}), the susceptibilities $\chi_j$ are interpolated at the series of values (\ref{eq:mh_n}) up to $n = 9$, which corresponds to $m_h(9) \simeq 3.9~\mbox{GeV} \simeq 0.75~m_b^{phys}$.

\begin{figure}[htb!]
\begin{center}
\includegraphics[scale=0.80]{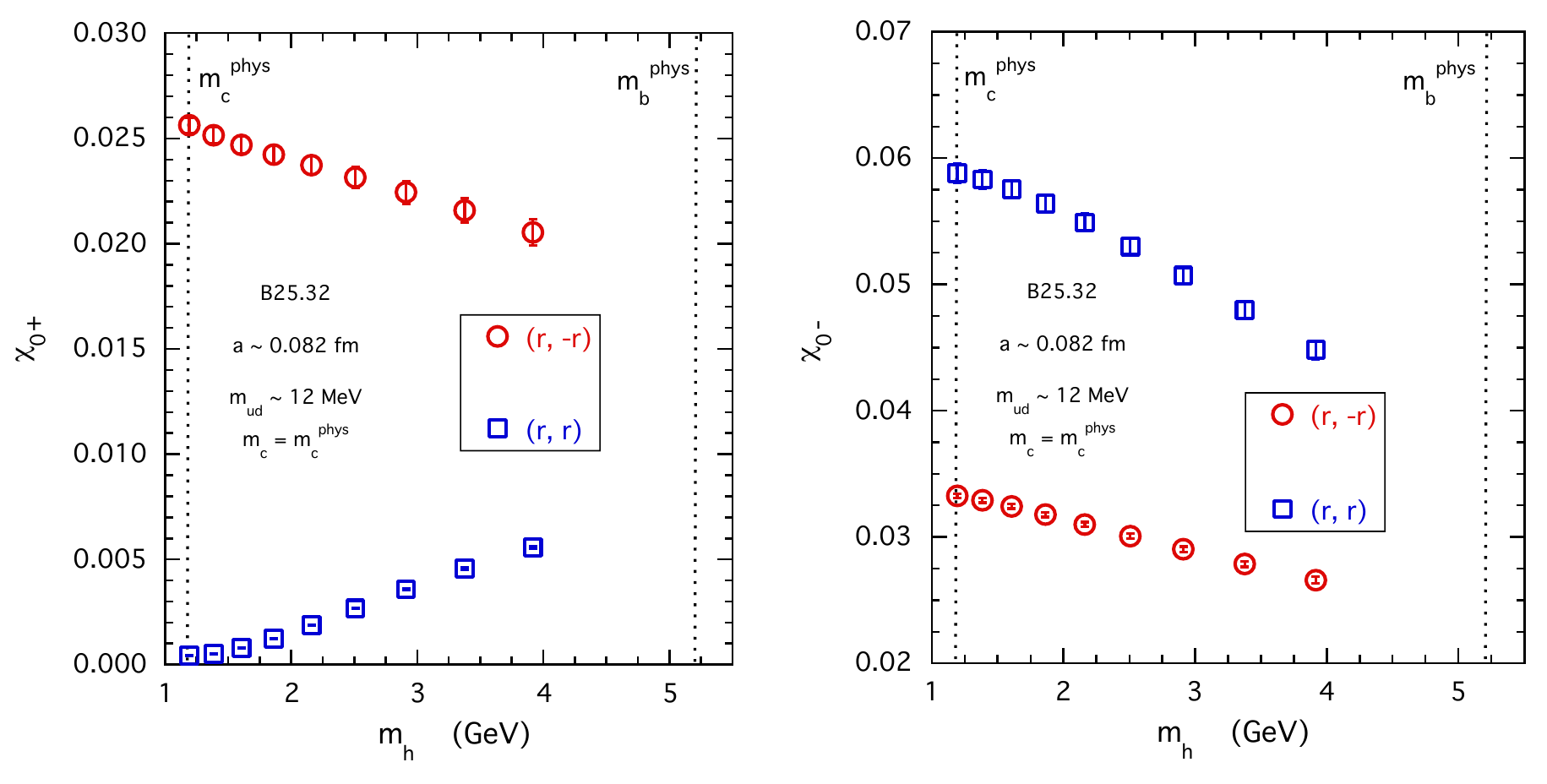}
\end{center}
\vspace{-1cm}
\caption{\it \small Vector and axial-vector longitudinal susceptibilities $\chi_{0^+}$ (left panel) and $\chi_{0^-}$ (right panel) corresponding to the gauge ensemble B25.32 and for the $h \to c$ transitions. The susceptibilities are obtained after a smooth interpolation at $m_c = m_c^{phys}$ and at the series of values of the heavy-quark mass $m_h$ given by Eq.~(\ref{eq:mh_n}) up to $n = 9$, i.e.~up to $m_h(9) \simeq 3.9$ GeV. The red circles correspond to the choice of opposite values $(r, -r)$ of the two valence-quark Wilson parameters, while the blue squares refer to the case of equal values $(r, r)$.}
\label{fig:VLAL_0}
%\end{figure}
%\begin{figure}[htb!]
\begin{center}
\includegraphics[scale=0.80]{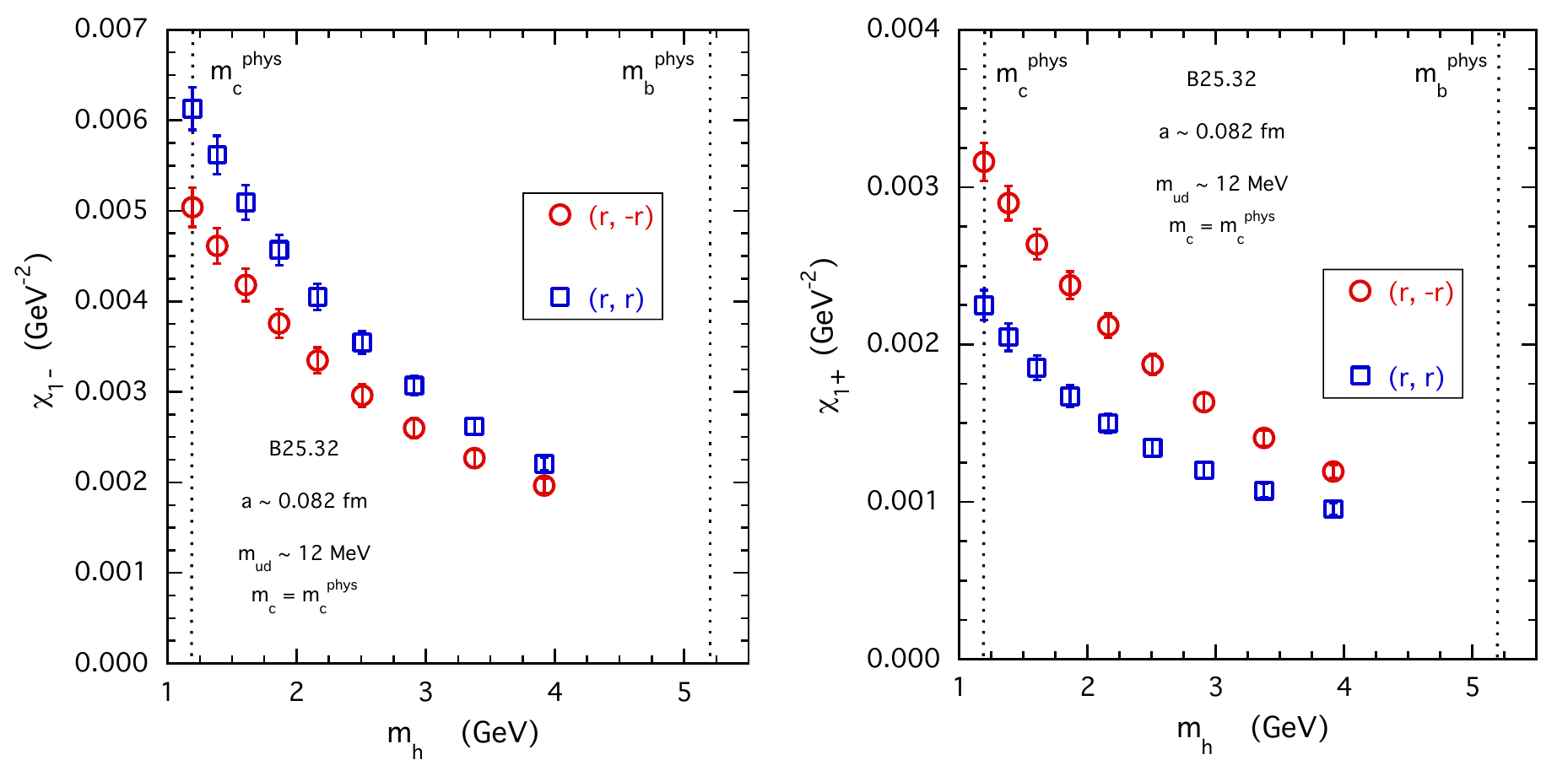}
\end{center}
\vspace{-1cm}
\caption{\it \small The same as in Fig.~\ref{fig:VLAL_0}, but for the vector and axial-vector transverse susceptibilities $\chi_{1^-}$ (left panel) and $\chi_{1^+}$ (right panel).}
\label{fig:VTAT_0}
\end{figure}

Using the gauge ensemble B25.32 as a representative case, our results for the vector and axial longitudinal susceptibilities are shown in Fig.~\ref{fig:VLAL_0} and  for the transverse ones in Fig.~\ref{fig:VTAT_0} at either opposite or equal values of the valence-quark Wilson parameters, which will be denoted hereafter by $(r, -r)$ and $(r, r)$.

The following comments are in order.
\begin{itemize}
\item Differences among results corresponding to the two $r$-combinations are expected to occur because of (twisted-mass) discretization effects, but it is clear that such differences are much larger in the longitudinal channels (particularly for $\chi_{0^+}$) with respect to the transverse cases.
\item For both $r$-combinations the transverse susceptibilities increase as $m_h$ increases, while the opposite behavior occurs for the longitudinal ones with the exception of $\chi_{0^+}$ in the $(r, r)$ case.
\item Because of charge conservation, the susceptibility $\chi_{0^+}$ evaluated for $m_h = m_c$ should vanish in the continuum limit. This seems to hold for the $(r, r)$ combination, while it is strongly violated in the $(r, -r)$ case.
\end{itemize}

The above observations point toward the presence of extra contributions coming from possible contact terms related to the product of two currents, which appear in all the correlators (\ref{eq:CVL12})-(\ref{eq:CP12}).
The issue of contact terms, which may affect the evaluation of the correlators for any lattice formulation of QCD, has been throughly investigated for our ETMC setups in Refs.~\cite{Burger:2014ada,Giusti:2017jof} in the case of the HVP contribution to the muon ($g-2$), which, as known, involves the product of two electromagnetic currents (i.e.~it corresponds to the degenerate case $m_h = m_c$). 
The main outcome is that contact terms may not vanish in the continuum limit due to the mixing of the product of two currents with terms proportional to second derivatives of the Dirac delta function.
A quick inspection of Eqs.~(\ref{eq:chiVL0})-(\ref{eq:chiAT0}) reveals that the longitudinal susceptibilities $\chi_{0^+}$ and $\chi_{0^-}$ are affected by contact terms (being second moments), while the transverse ones $\chi_{1^-}$ and $\chi_{1^+}$ are not (being fourth moments).

\begin{figure}[htb!]
\begin{center}
\includegraphics[scale=0.80]{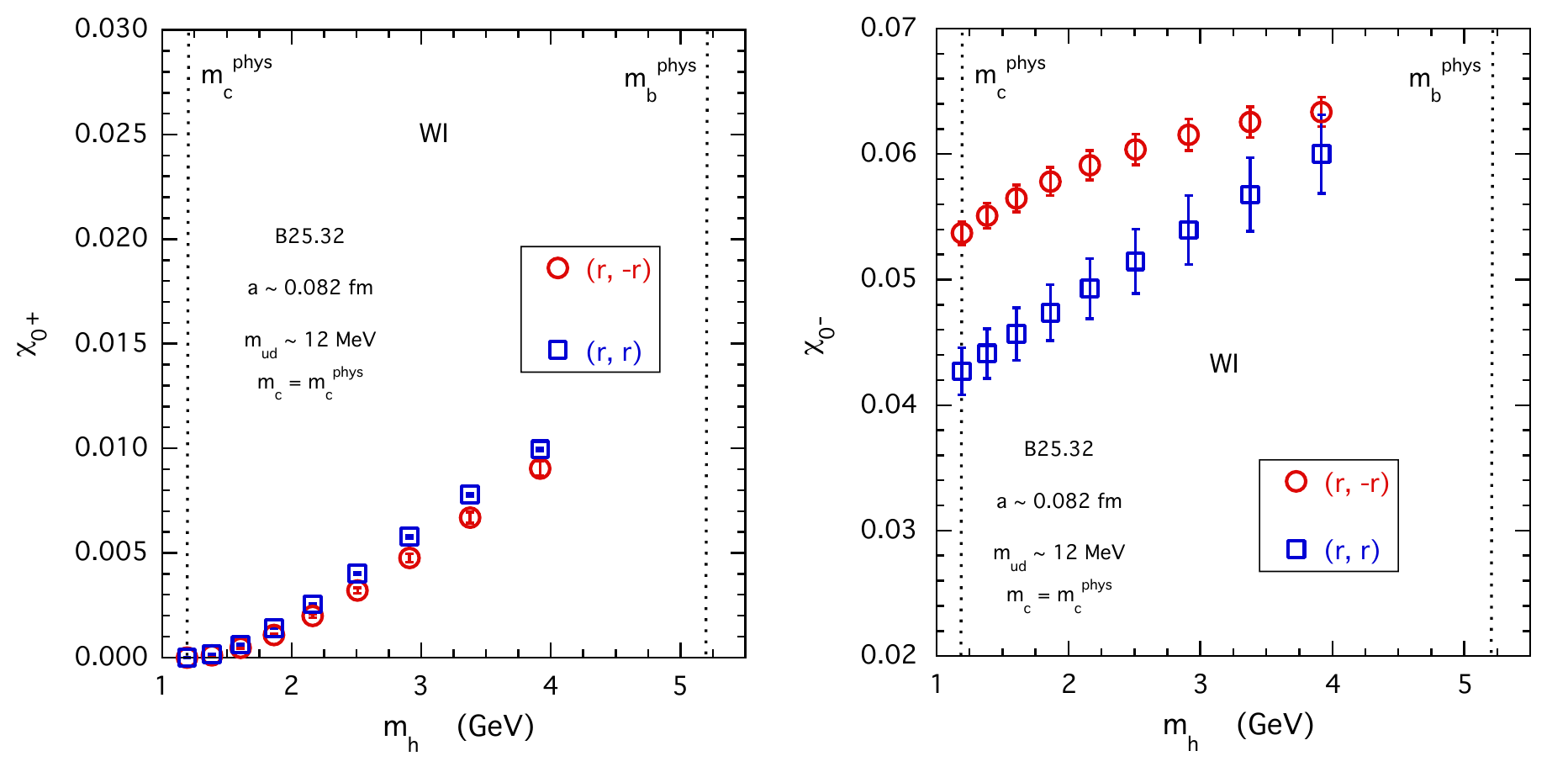}
\end{center}
\vspace{-1cm}
\caption{\it \small The same as in Fig.~\ref{fig:VLAL_0}, but using the definitions (\ref{eq:chiVL_WTI})-(\ref{eq:chiAL_WTI}) based on the use of the Ward-Takahashi identities. The vertical scales are kept the same as in Fig.~\ref{fig:VLAL_0}.}
\label{fig:VLAL_WI}
\end{figure}

A way to avoid contact terms is to replace Eqs.~(\ref{eq:chiVL0}) and (\ref{eq:chiAL0}) with the corresponding expressions (\ref{eq:chiVL_WI0}) and (\ref{eq:chiAL_WI0}) derived using WIs
\bea
     \label{eq:chiVL_WTI}
     \chi_{0^+} & ~ _{\overrightarrow{WI}} ~ & \frac{1}{12} (m_h - m_c)^2 \int_0^\infty dt ~ t^4 ~ C_S(t) ~ , ~ \\[2mm]
     \label{eq:chiAL_WTI}
     \chi_{0^-} & ~ _{\overrightarrow{WI}} ~ & \frac{1}{12} (m_h + m_c)^2 \int_0^\infty dt ~ t^4 ~ C_P(t) ~ , ~ 
\eea 
where $C_S(t)$ and $C_P(t)$ are given by Eqs.~(\ref{eq:CS12}) and (\ref{eq:CP12}), respectively.
In this way the longitudinal susceptibilities are evaluated using the fourth moments of the scalar and pseudoscalar correlators and they become essentially free from contact terms.
Note that in the degenerate case $m_h = m_c$ the susceptibility $\chi_{0^+}$ vanishes at any finite value of the lattice spacing.
The corresponding results for the longitudinal susceptibilities $\chi_{0^+}$ and $\chi_{0^-}$ in the case of the gauge ensemble B25.32 are shown in Fig.~\ref{fig:VLAL_WI}.

The comparison with Fig.~\ref{fig:VLAL_0} indicates very clearly that, without the effects of the contact terms, the differences among the results corresponding to the two $r$-combinations are much smaller.
More precisely, the contact terms heavily affect the vector longitudinal susceptibility $\chi_{0^+}$ for the combination $(r, -r)$, but they seem to be almost absent for the combination $(r, r)$.
Moreover, without contact terms both the vector and axial longitudinal susceptibilities increase as $m_h$ increases, as it happens for the transverse ones.

\subsection{Perturbative subtractions of contact terms and discretization effects}
\label{sec:contact}

Even if the use of the definitions (\ref{eq:chiVL_WTI})-(\ref{eq:chiAL_WTI}) are free from contact terms, it is very interesting to understand better the behaviour of the effects of the contact terms on the longitudinal susceptibilities (\ref{eq:chiVL0}) and (\ref{eq:chiAL0}).
In particular, we want to improve our understanding of the almost total absence of contact terms in the longitudinal vector susceptibility $\chi_{0^+}$ for the combination $(r, r)$ and of its huge quantitative impact for the other combination $(r, -r)$.

\begin{figure}[htb!]
\begin{center}
\includegraphics[scale=0.80]{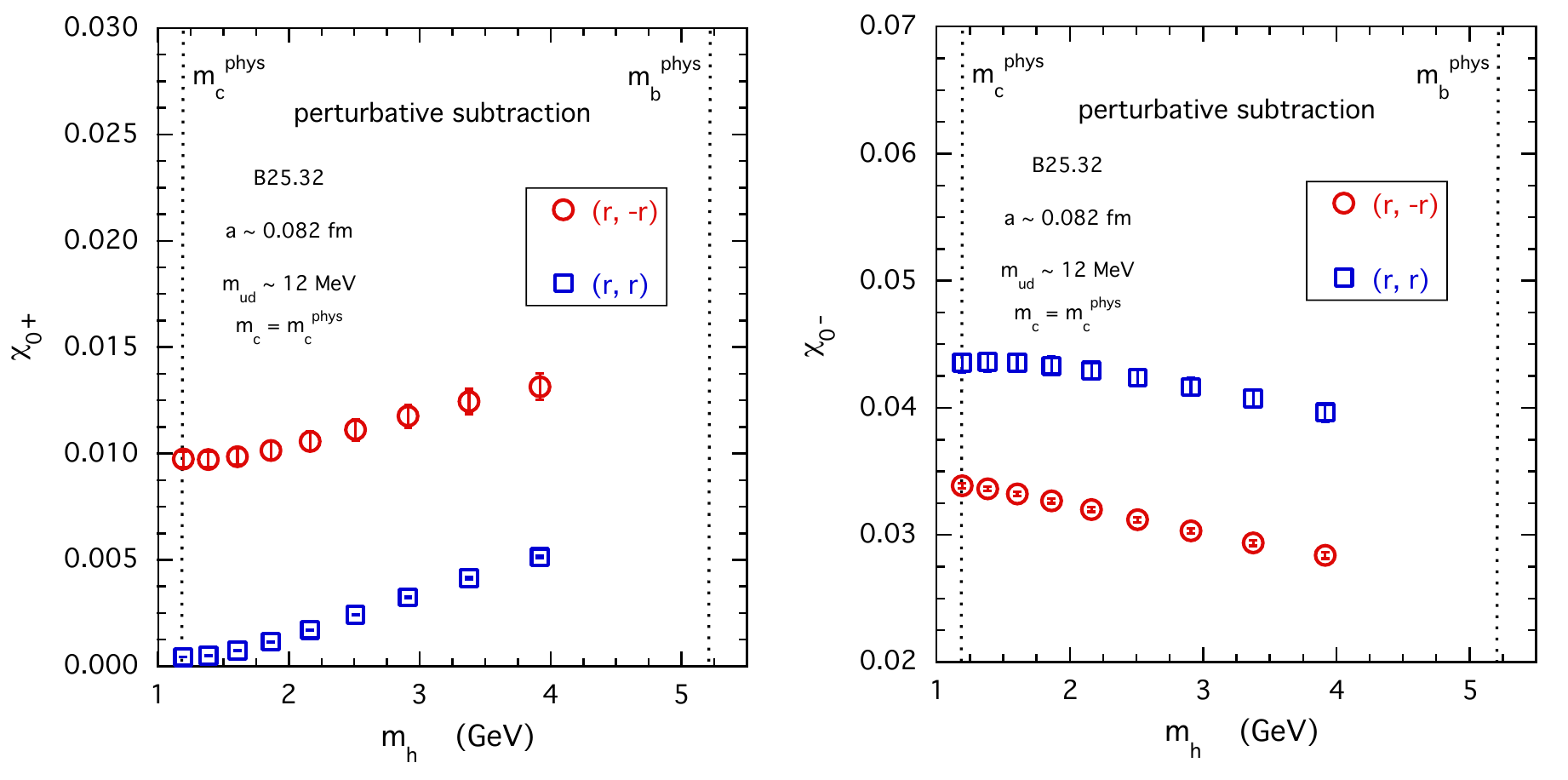}
\end{center}
\vspace{-1cm}
\caption{\it \small The same as in Fig.~\ref{fig:VLAL_0}, but after subtraction of the contact terms and of the discretization effects evaluated in the free theory, i.e.~at order ${\cal{O}}(\alpha_s^0)$, for our lattice setup [see Eq.~(\ref{eq:chij_pTsub})]. The vertical scales are kept the same as in Fig.~\ref{fig:VLAL_0}.}
\label{fig:VLAL_pTsub}
%\end{figure}
%\begin{figure}[htb!]
\begin{center}
\includegraphics[scale=0.80]{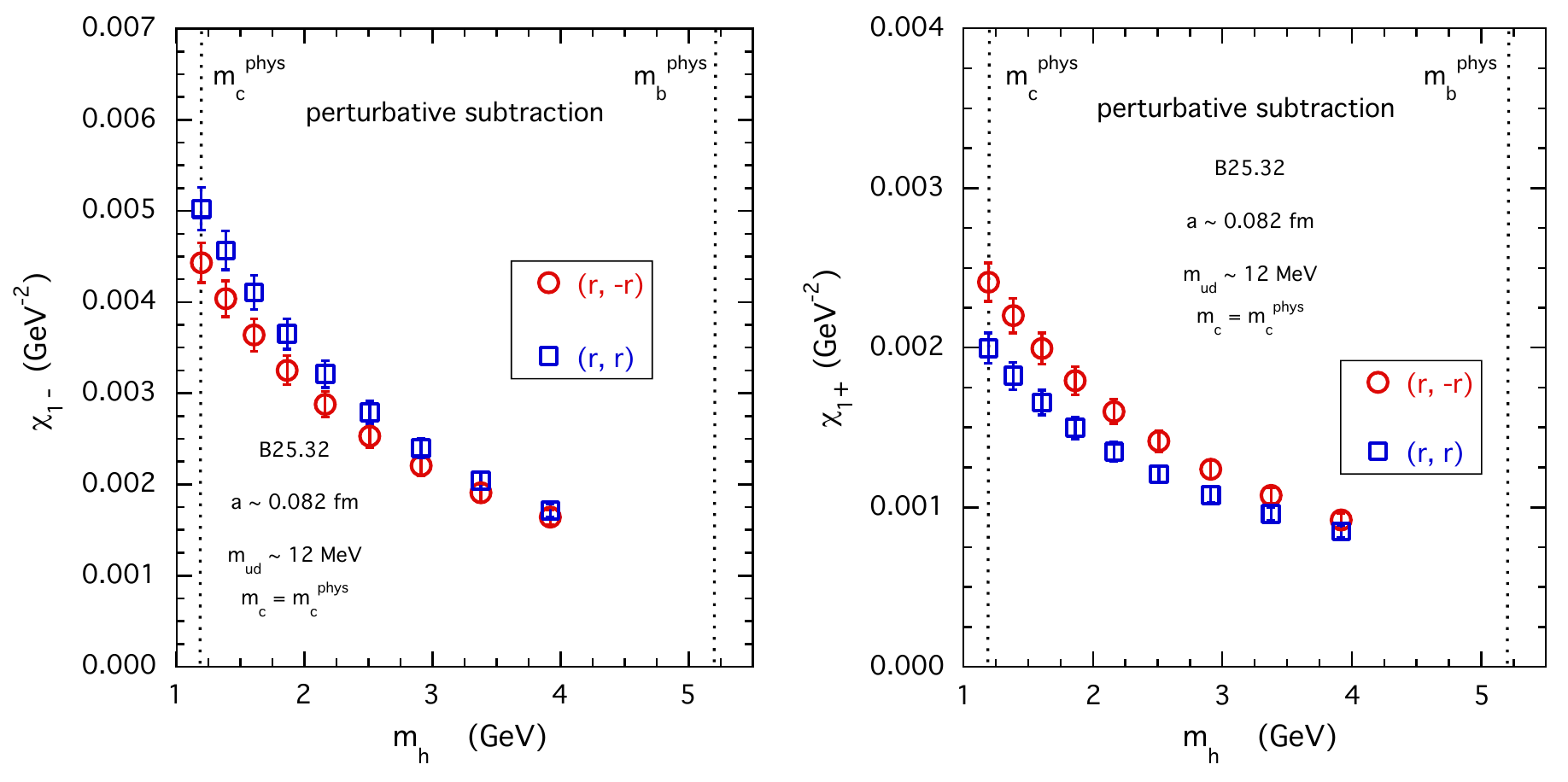}
\end{center}
\vspace{-1cm}
\caption{\it \small The same as in Fig.~\ref{fig:VTAT_0}, but after subtraction of the discretization effects evaluated in the free theory, i.e.~at order ${\cal{O}}(\alpha_s^0)$, for our lattice setup [see Eq.~(\ref{eq:chij_pTsub})]. The vertical scales are kept the same as in Fig.~\ref{fig:VTAT_0}.}
\label{fig:VTAT_pTsub}
\end{figure}

To this end, following the procedure discussed in Section VI of Ref.~\cite{DiCarlo:2021dzg}, we have calculated the susceptibilities $\chi_j$ ($j = \{ 0^+, 1^-, 0^-, 1^+ \}$) in the free theory on the lattice, i.e.~at order ${\cal{O}}(\alpha_s^0)$, using twisted-mass fermions with masses given in lattice units by $a m_h(n) = \lambda^{n-1} a m_c^{phys}$ for $n = 1, 2, ... ~$.
These results, which hereafter will be denoted by $\chi_j^{free}$, contain a physical contribution related to the LO term of PT, i.e.~at order ${\cal{O}}(\alpha_s^0)$, and the corrections due to contact terms and discretization effects present in the free theory for our lattice setup, i.e.~at all orders ${\cal{O}}(\alpha_s^0 a^m)$ with $m \geq 0$.
The former ones, which will be denoted by $\chi_j^{LO}$, are known from Ref.~\cite{Boyd:1997kz}, namely
\bea
      \label{eq:chiVL_LO}
      \chi_{0^+}^{LO} & = & \frac{1}{8 \pi^2 (1 - u^2)^3} \left[ (1 - u^2) (1 + u + u^2) (1 - 4u + u^2) - 6 u^3 \mbox{log}(u^2) \right] ~ , ~ \\[2mm]
      \label{eq:chiAL_LO}
      \chi_{0^-}^{LO} & = & \chi_{0^+}^{LO}(0)|_{u \to - u} ~ , ~ \\[4mm]
      \label{eq:chiVT_LO}
      m_h^2 ~\chi_{1-}^{LO} & = & \frac{1}{32 \pi^2(1 - u^2)^5} \left[ (1 - u^2) (3 + 4 u - 21 u^2 + 40 u^3 - 21 u^4 + 4 u^5 + 3 u^6)
                                                   \right. \nonumber \\[2mm]
                                         & + & \left.12 u^3 (2 - 3 u + 2 u^2) \mbox{log}(u^2) \right] ~ , ~ \\[2mm]
      \label{eq:chiAT_LO}
      \chi_{1^+}^{LO} & = & \chi_{1^-}^{LO}|_{u \to - u} ~ , ~     
\eea
where $u \equiv m_c / m_h \to \lambda^{1 - n}$.
Thus, we subtract from our non-perturbative susceptibilities $\chi_j$ the difference $\chi_j^{free} - \chi_j^{LO}$, namely
\be
    \label{eq:chij_pTsub}
    \chi_j \to \chi_j - \left[ \chi_j^{free} - \chi_j^{LO} \right] ~ . ~
\ee
The corresponding results are shown in Fig.~\ref{fig:VLAL_pTsub} for the longitudinal vector and axial susceptibilities and in Fig.~\ref{fig:VTAT_pTsub} for the transverse ones.

The perturbative calculation explains nicely that in the case of the longitudinal vector susceptibility $\chi_{0^+}$ the contact terms are almost absent for the combination $(r, r)$, while they show a huge quantitative effect for the other combination $(r, -r)$.
In the latter case the perturbative estimate is not enough to cancel out the contact term for degenerate quark masses $m_h = m_c^{phys}$, where $\chi_{0^+}$ is expected to vanish due to current conservation, and orders higher than ${\cal{O}}(\alpha_s^0)$ are clearly required.
By comparing the results shown in Fig.~\ref{fig:VLAL_pTsub} with those in Fig.~\ref{fig:VLAL_0} a reduction of the difference between the two $r$-combinations is visible also in the case of the longitudinal axial susceptibility $\chi_{0^-}$.

In the case of the transverse susceptibilities the contact terms are absent, but the subtraction of the discretization effects contained in the difference [$\chi_j^{free} - \chi_j^{LO}$] turns out to be beneficial.
By comparing the results shown in Fig.~\ref{fig:VTAT_pTsub} with those in Fig.~\ref{fig:VTAT_0} the difference between the combinations $(r, r)$ and $(r, -r)$ is significantly reduced. 

An alternative, rather effective, way to get rid of the contact terms in the susceptibility $\chi_{0^+}$ is to subtract the contact terms evaluated non-perturbatively at $m_h = m_c$, more precisely by using the formula 
\be
    \label{eq:CTsub}
     \chi_{0^+}(m_h, m_c) \to \chi_{0^+}(m_h, m_c) - \frac{1}{2} \left[ \chi_{0^+}(m_h, m_h) + \chi_{0^+}(m_c, m_c) \right] ~ , ~
\ee
obtaining in this way that $\overline{\chi}_{0^+}(m_h = m_c) = 0$ as in the case of the WI-based formula~(\ref{eq:chiVL_WTI}).

\begin{figure}[htb!]
\begin{center}
\includegraphics[scale=0.80]{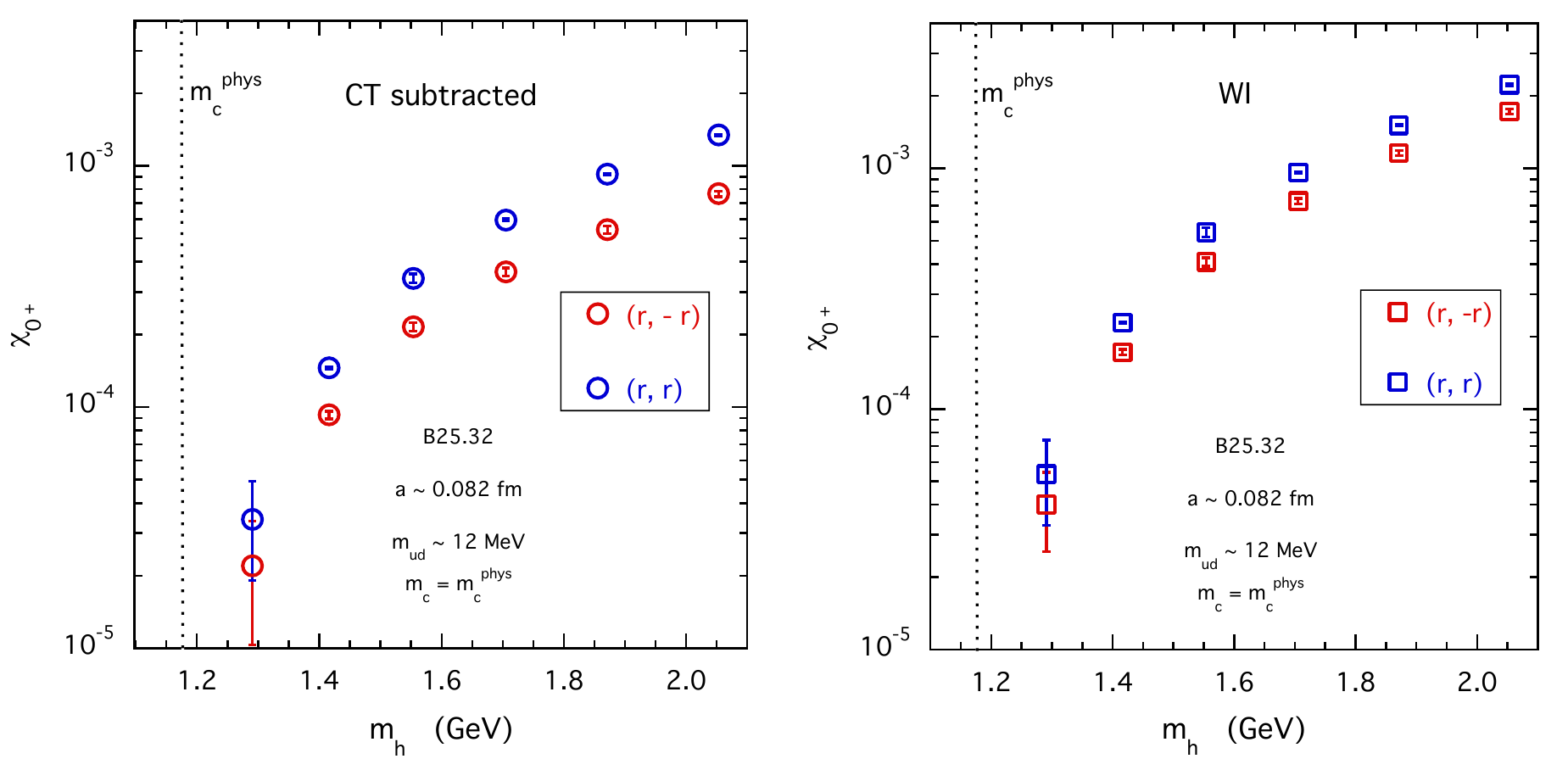}
\end{center}
\vspace{-1cm}
\caption{\it \small Vector longitudinal susceptibility $\chi_{0^+}$ obtained either using the subtraction procedure given by Eq.\,(\ref{eq:CTsub}) (left panel) or based on the WI~(\ref{eq:chiVL_WTI}) (right panel) corresponding to the gauge ensemble B25.32. \hspace*{\fill}} 
\label{fig:VL_comp}
\end{figure}

In Fig.\,\ref{fig:VL_comp} the results\footnote{Actually, since the calculations of $\chi_{0^+}(m_h, m_h)$ corresponding to the series of heavy-quark masses~(\ref{eq:mh_n}) are not available, what is illustrated in Fig.\,\ref{fig:VL_comp} has been obtained using the two-point correlation functions evaluated in Ref.~\cite{Carrasco:2013zta}.} obtained using Eq.\,(\ref{eq:CTsub}) are compared with those based on the WI formula~(\ref{eq:chiVL_WTI}).
A reassuring qualitative agreement is obtained, taking into account that discretization effects are different in the two procedures.
Note that smaller differences between the two $r$-combinations occur also in the case of the WI-based formula.

\begin{figure}[htb!]
\begin{center}
\includegraphics[scale=0.80]{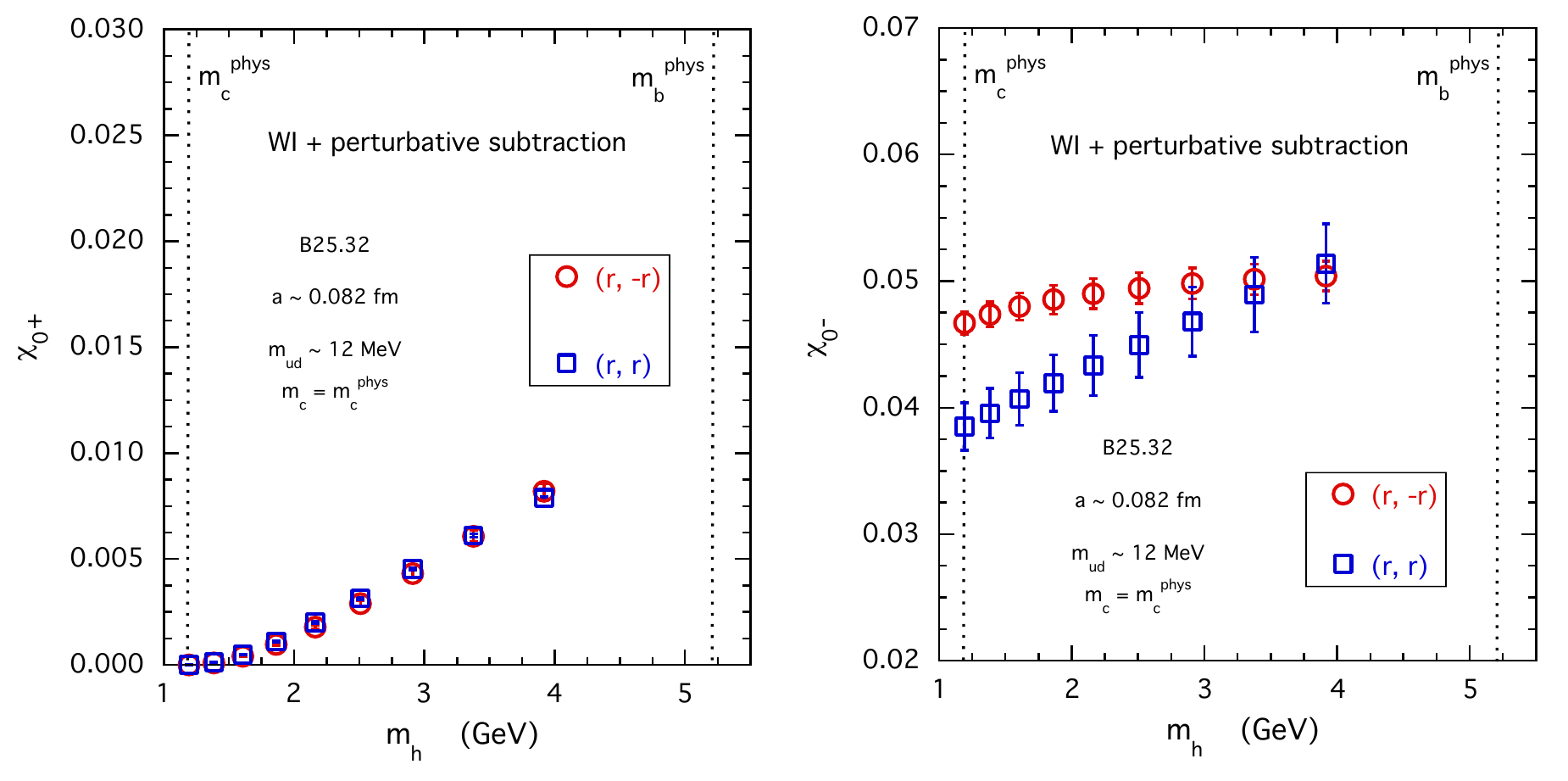}
\end{center}
\vspace{-1cm}
\caption{\it \small The same as in Fig.~\ref{fig:VLAL_WI}, but after the subtraction of the discretization effects evaluated in the free theory, i.e.~at order ${\cal{O}}(\alpha_s^0)$, for our lattice setup (see Eq.~(\ref{eq:chij_pTsub})). The vertical scales are kept the same as in Fig.~\ref{fig:VLAL_WI}.}
\label{fig:VLAL_WI_pTsub}
\end{figure}

Since the subtraction procedure given in Eq.\,(\ref{eq:CTsub}) is not applicable to the axial longitudinal susceptibility $\chi_{0^-}$, in what follows we only make use of the longitudinal susceptibilities based on the WIs, i.e.~on Eqs.~(\ref{eq:chiVL_WTI})-(\ref{eq:chiAL_WTI}), and shown in Fig.~\ref{fig:VLAL_WI_pTsub} after the subtraction of the discretization effects evaluated in the free theory according to Eq.~(\ref{eq:chij_pTsub}).

\section{The ETMC ratios}
\label{sec:ETMC_ratio}

According to the ETMC ratio method of Ref.~\cite{Blossier:2009hg} we now consider the ratios of the lattice susceptibilities $\overline{\chi}_j = \overline{\chi}_j[m_h(n); a^2, m_{ud}]$ interpolated for each ETMC gauge ensemble at subsequent values of the heavy-quark mass $m_h(n)$ given by Eq.~(\ref{eq:mh_n}), namely 
\be
    \label{eq:ETMC_ratios}
    R_j(n; a^2, m_{ud}) \equiv \frac{\chi_j[m_h(n); a^2, m_{ud}]}{\chi_j[m_h(n-1); a^2, m_{ud}]} ~ \frac{\rho_j[m_h(n)]}{\rho_j[m_h(n-1)]} ~ , ~
\ee
where $n = 2, 3, ... \, 9$ for $j = \{1^-, 0^-, 1^+ \}$ and $n = 3, 4, ... \, 9$ for $j = 0^+$ (because $\chi_{0^+}[m_c^{phys}; a^2, m_{ud}] = 0$ by charge conservation).

In Eq.~(\ref{eq:ETMC_ratios}) the factor $\rho_j(m_h)$ is introduced to guarantee that in the heavy-quark limit $m_h \to \infty$ (i.e., $n \to \infty$) one has $R_j \to 1$.
Using the PT results of Ref.~\cite{Boyd:1997kz} the above condition is satisfied by
\bea
    \label{eq:rhoL}
    \rho_{0^+} (m_h) = \rho_{0^-}(m_h) & = & 1 ~ , ~ \\[2mm]
    \label{eq:rhoT}
    \rho_{1^-}(m_h) = \rho_{1^+}(m_h) & = & (m_h^{pole})^2 ~ , ~ 
\eea
where $m_h^{pole}$ is the pole heavy-quark mass.
The latter one can be constructed from the $\overline{MS}(2~\rm{GeV})$ mass $m_h$ in two steps.
First, the PT scale is evolved from $\mu = 2$ GeV to the value $\mu = m_h$ using $\rm N^3LO$ perturbation theory \cite{Chetyrkin:1999pq} with four quark flavors ($n_\ell = 4$) and $\Lambda_{QCD}^{Nf = 4} = 294\,(12)$ MeV \cite{FLAG}, obtaining in this way $m_h(m_h)$.
Then, at order ${\cal{O}}(\alpha_s^2)$ the pole quark mass $m_h^{pole}$ is given in terms of the $\overline{MS}$ mass $m_h(m_h)$ by 
\bea
      m_h^{pole} & = & m_h(m_h) \left\{ 1 + \frac{4}{3} \frac{\alpha_s(m_h)}{\pi} + \left( \frac{\alpha_s(m_h)}{\pi} \right)^2 \right. \nonumber \\
                         & \cdot & \left. \left[ \frac{\beta_0}{24} (8 \pi^2 + 71) + \frac{35}{24} + \frac{\pi^2}{9} \mbox{ln}(2) - 
                                         \frac{7 \pi^2}{12} -\frac{\zeta_3}{6} \right] + {\cal{O}}(\alpha_s^3) \right\} ~ , 
       \label{eq:mh_pole}
\eea
where $\beta_0 = (33 - 2 n_\ell) / 12$ and $\zeta_3 \simeq 1.20206$.
The relation between $m_h^{pole}$ and $m_h(m_h)$ is known up to order ${\cal{O}}(\alpha_s^3)$ (see Refs.~\cite{Chetyrkin:1999qi,Melnikov:2000qh}), but the ratios of the transverse factors (\ref{eq:rhoT}) appearing in Eq.~(\ref{eq:ETMC_ratios}) turn out to be almost insensitive to such high-order corrections.

\begin{figure}[htb!]
\begin{center}
\includegraphics[scale=0.80]{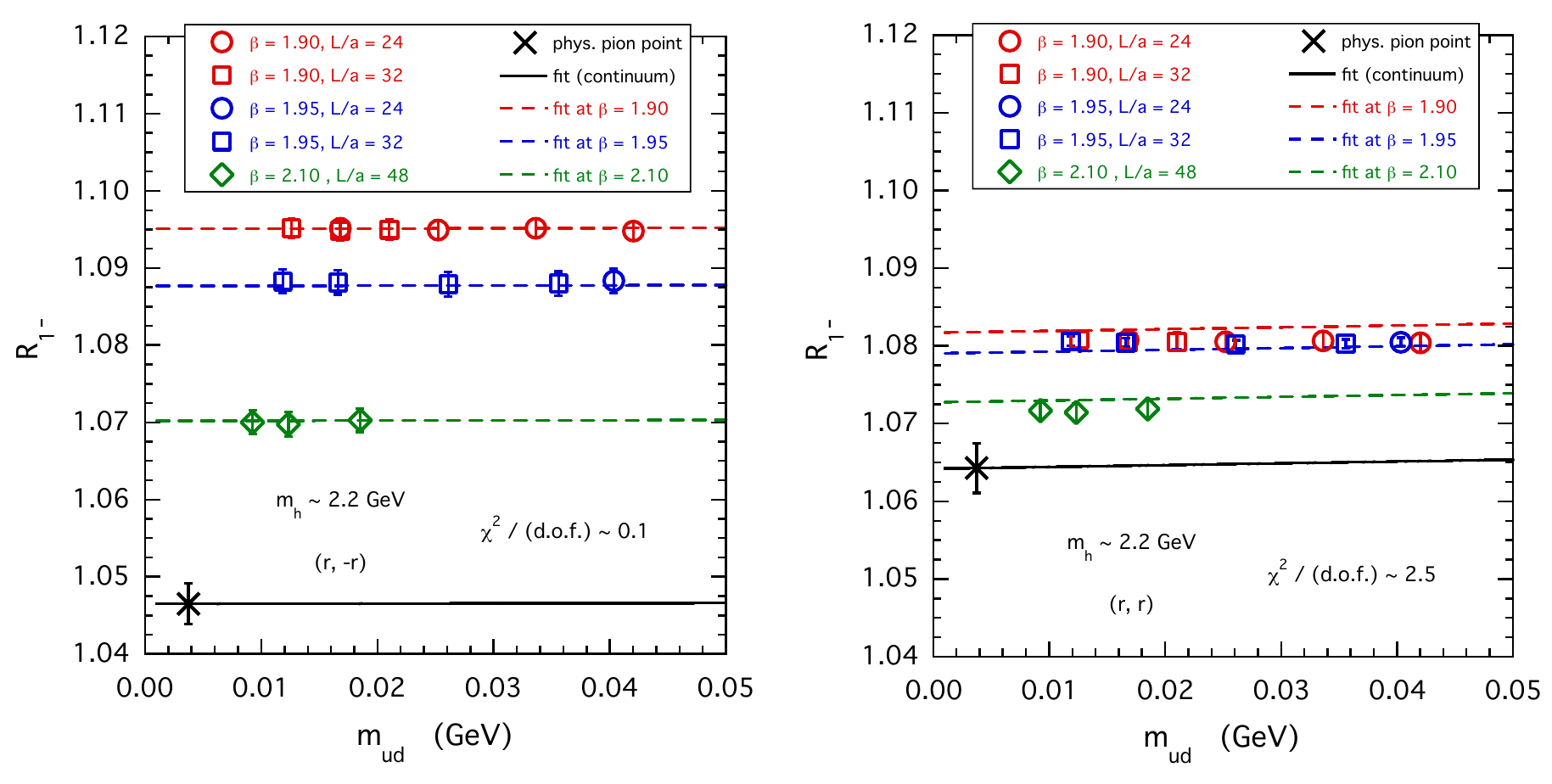}
\end{center}
\vspace{-1cm}
\caption{\it \small Light-quark mass dependence of the ratio of the susceptibilities corresponding to Eq.~(\ref{eq:ETMC_ratios}) for $j = 1^-$ and $n = 5$ for the two combinations $(r, -r)$ (left panel) and $(r,r)$ (right panel) of the Wilson $r$-parameters. The solid and dashed lines represent the results of the fitting function (\ref{eq:ratiofit_nopriors}) evaluated in the continuum limit and at each value of $\beta$. The crosses represent the value of the ratio extrapolated at the physical pion point ($m_{ud} = m_{ud}^{phys}$) and in the continuum limit.}
\label{fig:RVT_nopriors}
\end{figure}

Thanks to the large correlation between the numerator and the denominator in Eq.~(\ref{eq:ETMC_ratios}) the statistical uncertainty of the ETMC ratios $R_j(n; a^2, m_{ud})$ is much smaller than those of the separate susceptibilities and it may reach the permille level, as it is illustrated in Fig.~\ref{fig:RVT_nopriors} in the case $j = 1^-$ and $n=5$.
Moreover, the light-quark mass dependence of the ratios is very mild, because it comes entirely from the light sea quarks.
Therefore, for each value of the heavy-quark mass $m_h(n)$ we fit the lattice data by adopting a simple linear Ansatz in the light-quark mass $m_{ud}$  as well as in the values of the squared lattice spacing $a^2$, since in our lattice setup the susceptibilities are ${\cal{O}}(a)$-improved, namely
\be
    \label{eq:ratiofit_nopriors}
    R_j(n; a^2, m_{ud}) = R_j(n) \left[ 1 + A_1 \left( m_{ud} - m_{ud}^{phys} \right) + D_1 ~ \frac{a^2}{r_0^2} \right] ~ , ~
\ee
where  $r_0 \simeq 0.47$ fm is the value of the Sommer parameter (determined for our lattice setup in Ref.~\cite{Carrasco:2014cwa}) and $R_j(n)$ stands for $R_j(n; 0, m_{ud}^{phys})$. For sake of simplicity, we have dropped in the notation of the coefficients $A_1$ and $D_1$ their dependence on the specific channel $j$ (as well as on the specific value of the heavy-quark mass $m_h$).

\begin{figure}[htb!]
\begin{center}
\includegraphics[scale=0.80]{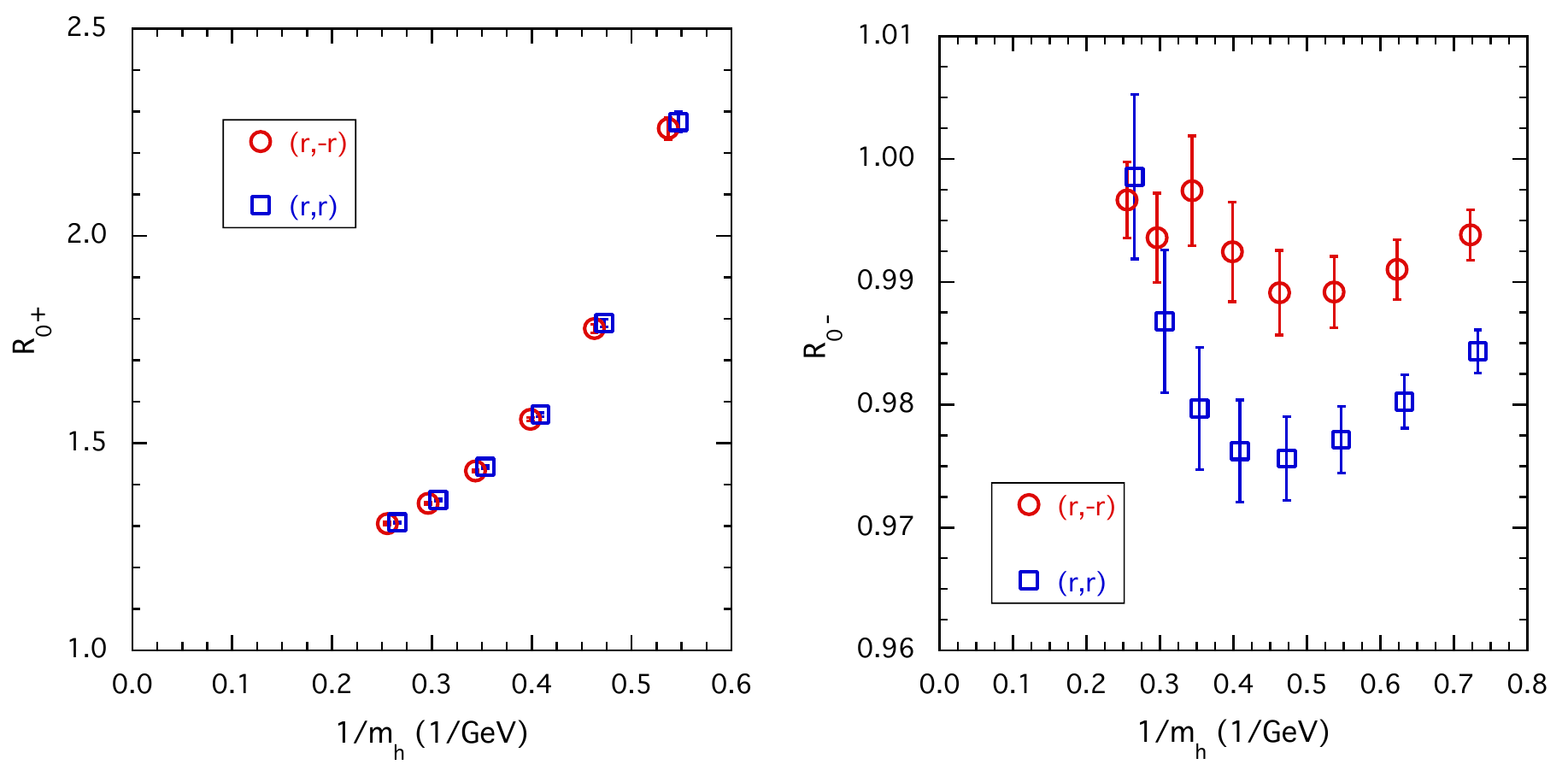}
\end{center}
\vspace{-1cm}
\caption{\it \small The longitudinal susceptibility ratios $R_{0^+}$ (left panel) and $R_{0^-}$ (right panel), after extrapolation to the physical pion point and to the continuum limit according to Eq.~(\ref{eq:ratiofit_nopriors}), versus the inverse heavy-quark mass $1 / m_h$ for the two combinations $(r, -r)$ and $(r,r)$ of the Wilson $r$-parameters. The values of $n$ in Eq.~(\ref{eq:ratiofit_nopriors}) range from $3$ to $9$ for $R_{0^+}$ and from $2$ to $9$ for $R_{0^-}$. The locations of the results for the combination $(r, r)$ are slightly moved to the right for a better reading.}
\label{fig:RVLAL_nopriors}
%\end{figure}
%\begin{figure}[htb!]
\begin{center}
\includegraphics[scale=0.80]{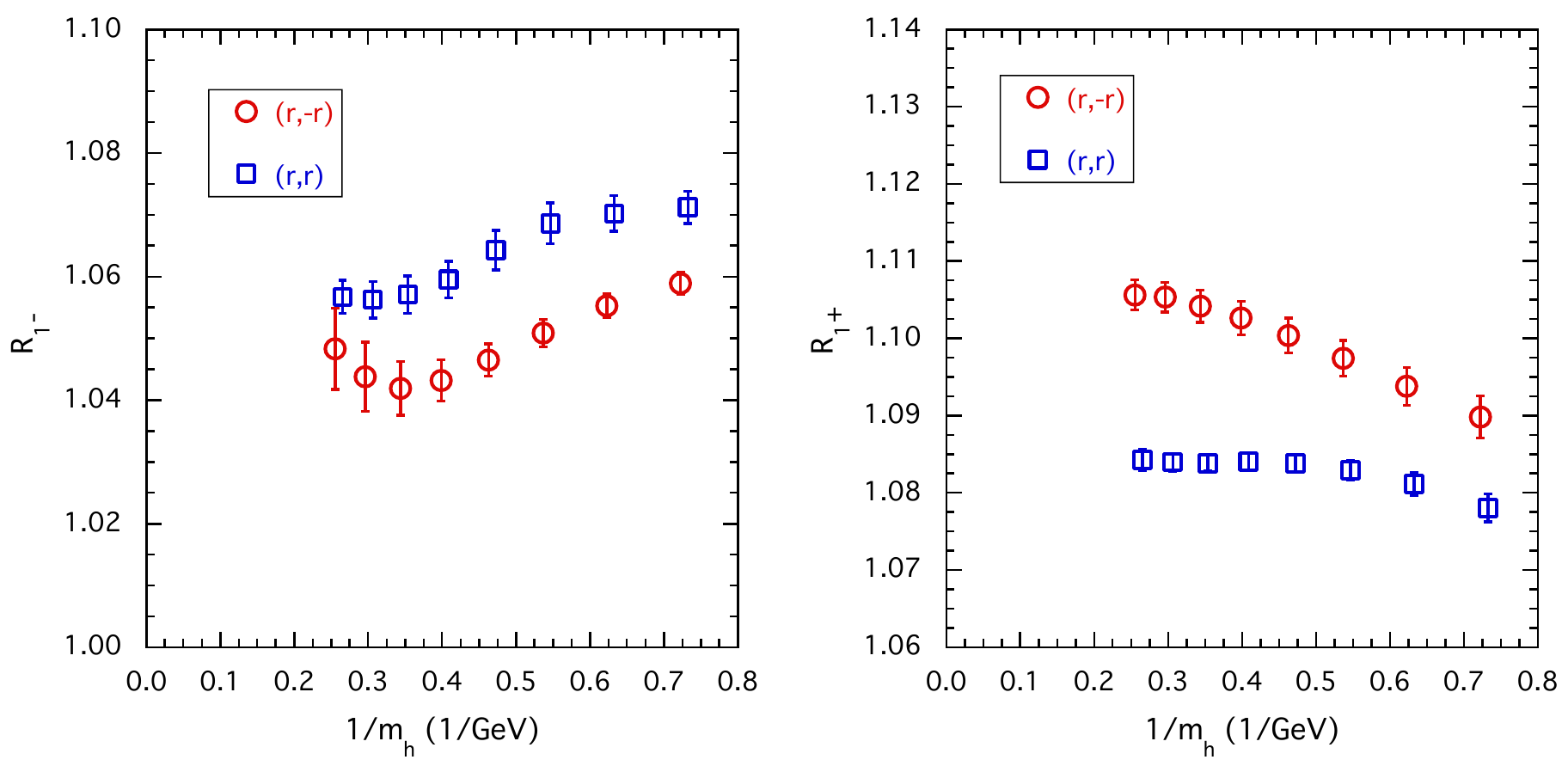}
\end{center}
\vspace{-1cm}
\caption{\it \small The same as in Fig.~\ref{fig:RVLAL_nopriors}, but in the case of the transverse susceptibility ratios $R_{1^-}$ (left panel) and $R_{1^+}$ (right panel) after extrapolation to the physical pion point and to the continuum limit according to Eq.~(\ref{eq:ratiofit_nopriors}).}
\label{fig:RVTAT_nopriors}
\end{figure}

The results obtained with the fitting function (\ref{eq:ratiofit_nopriors}) are shown in Fig.~\ref{fig:RVT_nopriors} as an illustrative example.
The quality of the fitting procedure may be quite good in several cases, as shown in the left panel of  Fig.~\ref{fig:RVT_nopriors} where the value of $\chi^2/(d.o.f.)$ is significantly less than $1$, but it may be also quite poor, as shown in the right panel of  Fig.~\ref{fig:RVT_nopriors} where the value of $\chi^2/(d.o.f.)$ is significantly larger than $1$.
In the latter cases discretization effects beyond the order ${\cal{O}}(a^2)$ seem to be required.
Moreover, in Eq.~(\ref{eq:ratiofit_nopriors}) the coefficient $R_j(n)$ represents the value of the ETMC ratio extrapolated to the physical pion point and to the continuum limit.
However, the susceptibilities corresponding to the two combinations $(r, -r)$ and $(r,r)$ of the Wilson $r$-parameters should differ only by discretization effects (at least of order ${\cal{O}}(a^2)$ in our maximally twisted setup).
This means that the value of $R_j(n)$ should be independent of the choice of the Wilson $r$-parameters.
This is not the case (except for $j = 0^+$) as it is clearly shown in Figs.~\ref{fig:RVLAL_nopriors} and \ref{fig:RVTAT_nopriors}.
The conclusion is that the Anstaz (\ref{eq:ratiofit_nopriors}) is not sufficient for describing the lattice data, since discretization effects beyond the order ${\cal{O}}(a^2)$ should be taken into account.

A possible option is to extend the calculation of the subtraction term $[\chi_j^{free} -  \chi_j^{LO}]$ (see Eq.~(\ref{eq:chij_pTsub})) to the NLO in perturbation theory, i.e.~at order ${\cal{O}}(\alpha_s)$, in order to eliminate at least discretization effects of order ${\cal{O}}(\alpha_s a^m)$ with $m > 0$.
While waiting for the above two-loop calculations, which will be addressed in a future work, an alternative option is to include explicitly in the fitting Ansatz a term proportional to $a^4$.
Since our lattice setup includes only three values of the lattice spacing, this procedure requires the use of a prior.
Thus, with respect to Eq.~(\ref{eq:ratiofit_nopriors}) we add a term proportional to $a^4$, namely
\be
    \label{eq:ratiofit}
    R_j(n; a^2, m_{ud}) = R_j(n) \left[ 1 + A_1 \left( m_{ud} - m_{ud}^{phys} \right) + D_1 ~ \frac{a^2}{r_0^2} + D_2 ~ \frac{a^4}{r_0^4} \right] ~ , ~
\ee
and we include a (gaussian) prior, $D_{prior}$, on the two parameters $D_1$ and $D_2$ by increasing the $\chi^2$ variable with the additive term $(D_1 / D_{prior})^2 + (D_2 / D_{prior}^2)^2$.
We remind that, for sake of simplicity, we have dropped in the parameters $A_1$, $D_1$ and $D_2$ their explicit dependence on the specific channel $j$ as well as on the specific value of the heavy-quark mass $m_h$. As for the dependence of the discretization terms on $m_h$, we have found that the fitting values of $D_1$ and $D_2$ generally increase as $m_h$ increases and they are roughly proportional to $m_h$. The quality of the fits based on Eq.~(\ref{eq:ratiofit}) is quite good (namely $\chi^2/(d.o.f.) \lesssim 0.1$) for all channels and heavy-quark masses.

It turns out that by increasing $D_{prior}$ the difference between the values $R_j(n)$ corresponding to the two combinations $(r, -r)$ and $(r,r)$ of the Wilson $r$-parameters decreases at the price of increasing the uncertainty of $R_j(n)$.
For $D_{prior} \simeq 20$ the above difference becomes much smaller than the uncertainty.
Any further increase of $D_{prior}$ does not lead to any improvement and produces only an increase of the uncertainty of $R_j(n)$\footnote{In principle, $D_{prior}$ can be chosen to be different for different values of the heavy-quark mass $m_h$ and for different channels $j$. 
In practice we have adopted a unique value of $D_{prior} $ common to all heavy-quark masses and channels.}.
The value obtained for $D_{prior}$ corresponds approximately to $r_0^2 / a_{coarse}^2$, where for our lattice setup $a_{coarse} \approx 0.1$ fm.
Indeed, as it can be seen from the illustrative case of  the right panel of Fig.~\ref{fig:RVT_nopriors}, the discretization effects flatten significantly at the coarsest lattices with respect to a pure $a^2$-scaling. This can be realized when $D_2 ~ a_{coarse}^2 / r_0^2 \approx - D_1$. Putting $|D_1| \approx D_{prior}$ and $|D_2| \approx D_{prior}^2$, it follows that $D_{prior} \approx r_0^2 / a_{coarse}^2$. Later in Section~\ref{sec:alternative} we perform an alternative extrapolation to the continuum limit (see Eq.~(\ref{eq:combinedfit})), whose results are quite reassuring about our good control of the extrapolation model (see Figs.~\ref{fig:RVLAL_combined} and \ref{fig:RVTAT_combined}).

\begin{figure}[htb!]
\begin{center}
\includegraphics[scale=0.80]{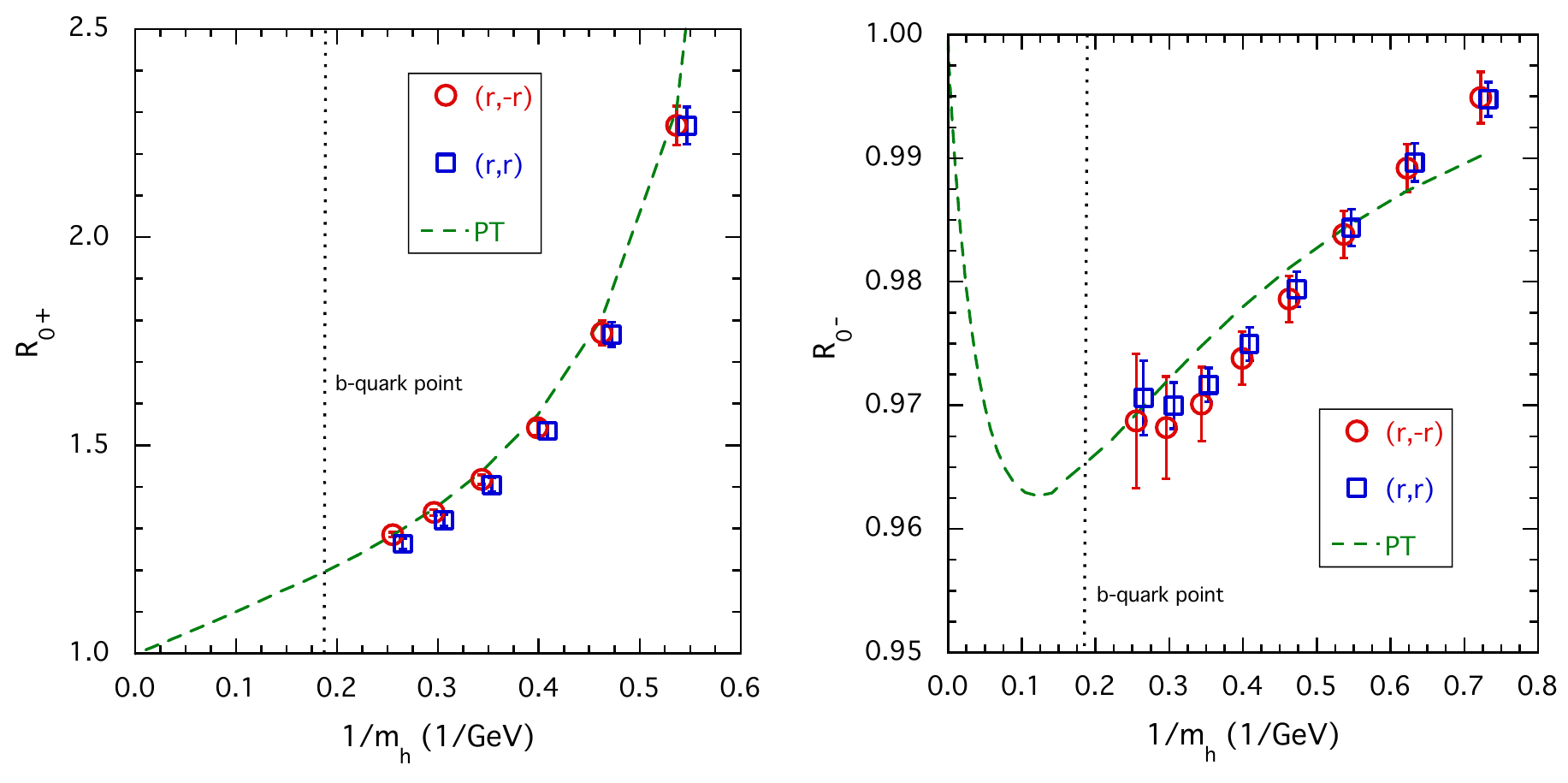}
\end{center}
\vspace{-1cm}
\caption{\it \small The same as in Fig.~\ref{fig:RVLAL_nopriors} in the case of the longitudinal susceptibility ratios $R_{0^+}$ (left panel) and $R_{0^-}$ (right panel), but using the fitting function (\ref{eq:ratiofit}) and adopting the value $D_{prior} = 20$. The dashed lines represent the PT predictions at order ${\cal{O}}(\alpha_s)$ from Ref.~\cite{Boyd:1997kz} (see Eq.~(\ref{eq:ratios_PT})). The vertical dotted line indicates the location of the physical $b$-quark point.}
\label{fig:RVLAL}
%\end{figure}
%\begin{figure}[htb!]
\begin{center}
\includegraphics[scale=0.80]{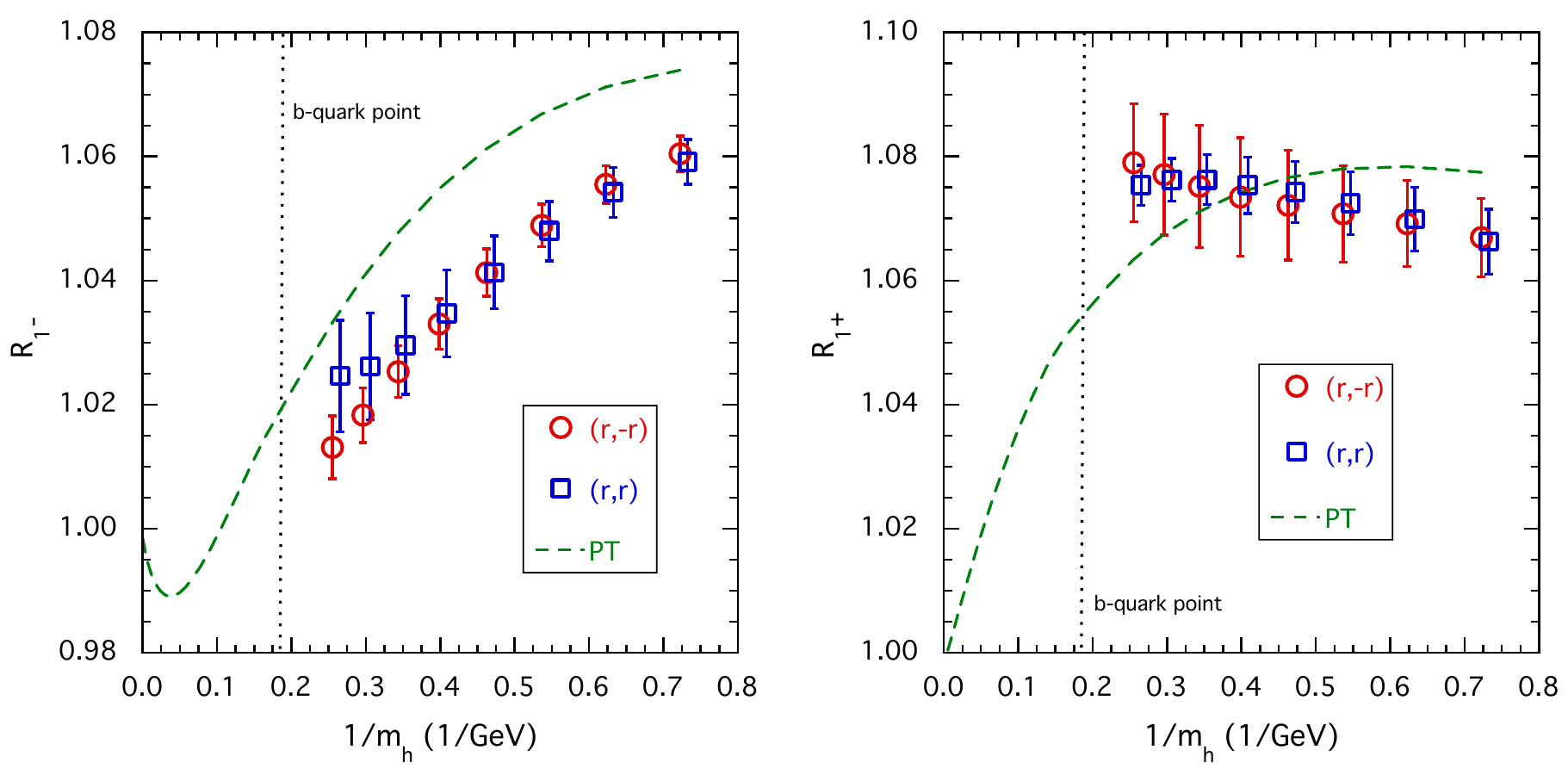}
\end{center}
\vspace{-1cm}
\caption{\it \small The same as in Fig.~\ref{fig:RVLAL}, but in the case of the transverse susceptibility ratios $R_{1^-}$ (left panel) and $R_{1^+}$ (right panel).}
\label{fig:RVTAT}
\end{figure}

The results obtained with the fitting function~(\ref{eq:ratiofit}) and adopting $D_{prior} = 20$ are shown in Figs.~\ref{fig:RVLAL} and \ref{fig:RVTAT}.
They are compared with the PT predictions at order ${\cal{O}}(\alpha_s)$ based on the findings of Ref.~\cite{Boyd:1997kz}, namely for $j = \{ 0^+, 1^-, 0^-, 1^+ \}$
\be
     \label{eq:ratios_PT}
     R_j^{PT}(n) \equiv \frac{\chi_j^{PT}[m_h^{pole}(n)]}{\chi_j^{PT}[m_h^{pole}(n-1)]} ~ \frac{\rho_j[m_h^{pole}(n)]}{\rho_j[m_h^{pole}(n-1)]} ~ , ~
\ee
where $\rho_j$ is defined by Eqs.~(\ref{eq:rhoL})-(\ref{eq:rhoT}) and 
\be
    \label{eq:chi_PT}
     \chi_j^{PT} = \chi_j^{LO} + \frac{\alpha_s(m_h^{pole})}{\pi} \chi_j^{NLO}
\ee
with $\chi_j^{LO}$ given by Eqs.~(\ref{eq:chiVL_LO})-(\ref{eq:chiAT_LO}) and
\bea
     \label{eq:chiVL_NLO}
     \chi_{0^+}^{NLO} & = & \frac{1}{48 \pi^2 (1 - u^2)^4} \left\{ (1 - u^2)^2 [1 - 36 u - 22 u^2 - 36 u^3 + u^4] - 2u (1 - u^2) \mbox{ln}(u^2) \right. \nonumber \\[2mm]
                                 & \cdot & \left.  [9 + 4 u + 66 u^2 + 4 u^3 + 9 u^4] - 4 u^3 \mbox{ln}^2(u^2) [9 + 18 u^2 - 2 u^3 - 3 u^4 + u^5] \right. \nonumber \\[2mm] 
                                 & + & \left. 8 (1 - u^2)^3 ~ \mbox{Li}_2(1 - u^2) [1 - 3 u + u^2] \right\}~ , ~ \\[4mm]
     \label{eq:chiAL_NLO}
     \chi_{0^-}^{NLO} & = &  \chi_{0^+}^{NLO}|_{u \to - u} ~ , ~
\eea
{\small
\bea
     \label{eq:chiVT_NLO}
     (m_h^{pole})^2 ~ \chi_{1^-}^{NLO} & = & \frac{1}{576 \pi^2 (1 - u^2)^6} \left\{ (1 - u^2)^2 [75 + 360 u - 1031 u^2 + 1776 u^3 -1031 u^4 + 
                                                                      360 u^5 \right. \nonumber \\[2mm]
                                & + & \left. 75 u^6] + 4u (1 - u^2) \mbox{ln}(u^2) [18 - 99 u + 732 u^2 - 1010 u^3 + 732 u^4 + 99 u^5 \right. \nonumber \\[2mm] 
                                & - & \left. 18 u^6] + 4 u^3 \mbox{ln}^2(u^2) [108 - 324 u + 648 u^2 - 456 u^3 + 132 u^4 + 59 u^5 - 12 u^6 \right. \nonumber \\[2mm] 
                                & - & \left. 9 u^7] + 8 (1 - u^2)^3 ~ \mbox{Li}_2(1 - u^2) [9 + 12 u - 32 u^2 + 12 u^3 + 9 u^4] \right\} ~ , ~ \\[4mm]
     \label{eq:chiAT_NLO}
     \chi_{1^+}^{NLO} & = & \chi_{1^-}^{NLO}|_{u \to - u} ~ . ~
\eea
}In Eqs.~(\ref{eq:chiVL_NLO})-(\ref{eq:chiAT_NLO}) the dilogarithm is defined as $ \mbox{Li}_2(z) = - \int_0^z dz^\prime \mbox{ln}(1 - z^\prime) / z^\prime$ and the variable $u$ is given by $u = m_c^{pole} / m_h^{pole}$.

From Figs.~\ref{fig:RVLAL} and \ref{fig:RVTAT} it can be seen that non-perturbative effects as well as PT ones at order ${\cal{O}}(\alpha_s^m)$ with $m \geq 2$ appear to be at the percent level on both the longitudinal and the transverse ratios.

\subsection{Extrapolation to the $b$-quark point}
\label{sec:bquark}

\begin{figure}[htb!]
\begin{center}
\includegraphics[scale=0.80]{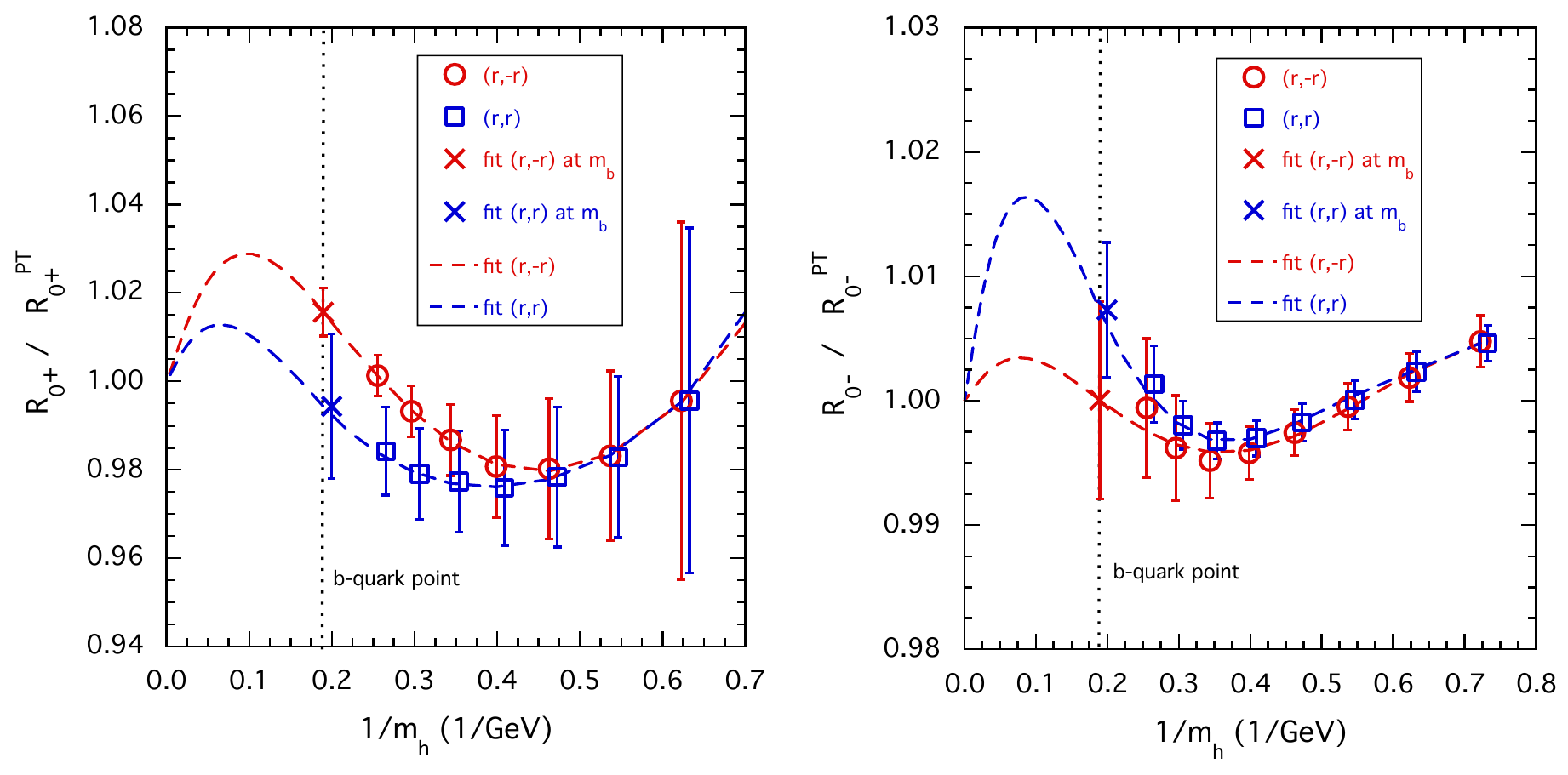}
\end{center}
\vspace{-1cm}
\caption{\it \small Longitudinal double ratios $R_{0^+} / R_{0^+}^{PT}$ (left panel) and $R_{0^-} / R_{0^-}^{PT}$ (right panel) versus the inverse heavy-quark mass $1 / m_h$. The dashed lines represent the results obtained with the fitting function (\ref{eq:doubleratio_fit}), while the crosses are the fit values at the physical $b$-quark point, indicated by vertical dotted lines. We remind that the locations of the results for the combination $(r, r)$ are slightly moved to the right for a better reading.}
\label{fig:DRVLAL}
%\end{figure}
%\begin{figure}[htb!]
\begin{center}
\includegraphics[scale=0.80]{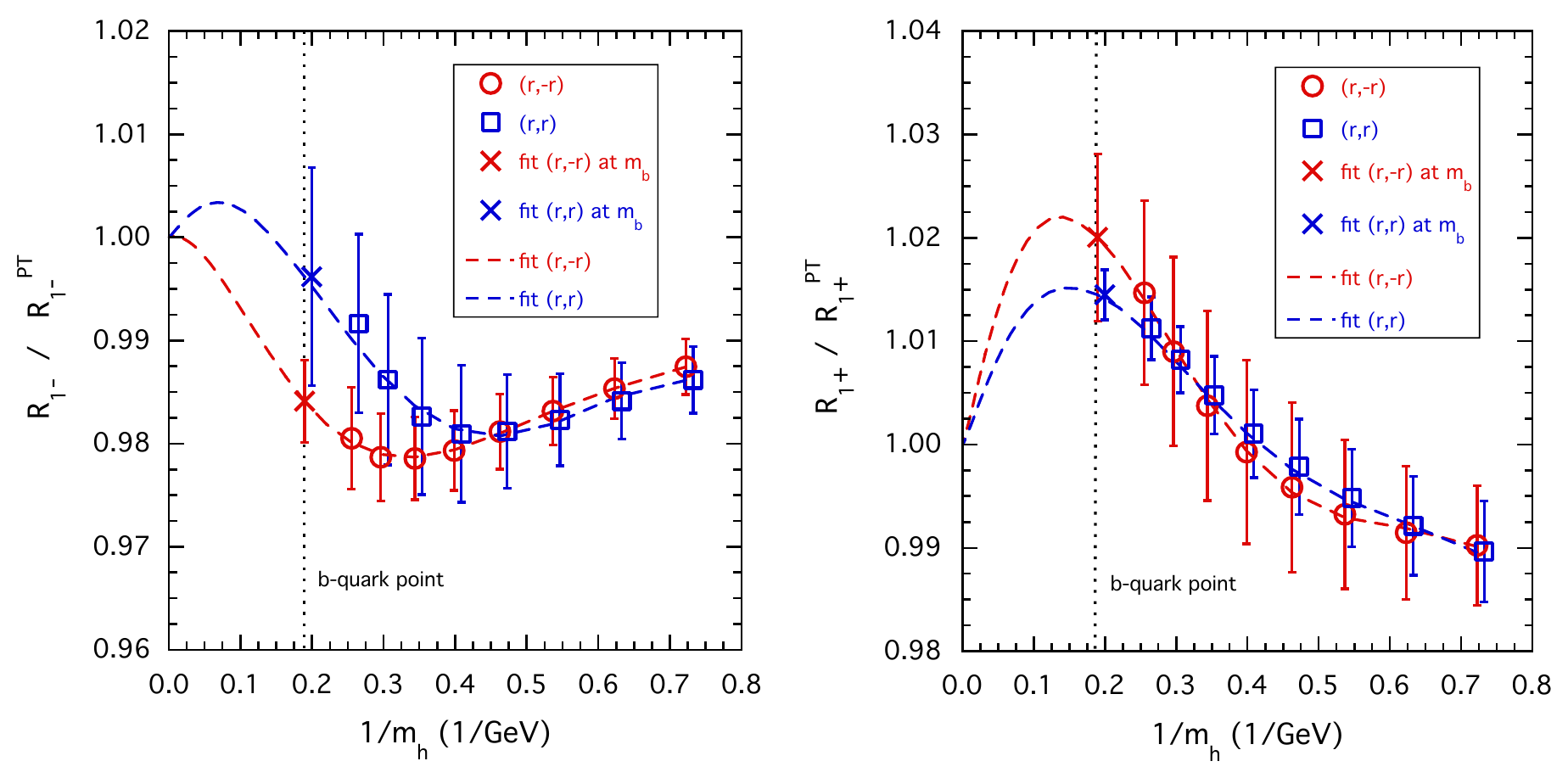}
\end{center}
\vspace{-1cm}
\caption{\it \small The same as in Fig.~\ref{fig:DRVLAL}, but in the case of the transverse double ratios $R_{1^-} / R_{1^-}^{PT}$ (left panel) and $R_{1^+} / R_{1^+}^{PT}$ (right panel).}
\label{fig:DRVTAT}
\end{figure}

The important feature of the ETMC ratio method is that the extrapolation to the physical $b$-quark point of the ratios $R_j$ can be carried out taking advantage of the fact that by construction
\be
     \mbox{lim}_{n \to \infty} ~ R_j(n) = 1 ~ . ~
\ee
However, in order to make smoother the extrapolation to the physical $b$-quark point (which we remind corresponds to $n = 11$ for our choice of $\lambda$) we consider the double ratios $R_j(n) / R_j^{PT}(n)$.
In this way the nontrivial heavy-quark mass dependence of $R_j^{PT}$ is taken into account and the deviations from unity are expected to be further reduced, as it is shown in Figs.~\ref{fig:DRVLAL}-\ref{fig:DRVTAT}.

Since the PT contribution $R_j^{PT}$ is given up to order ${\cal{O}}(\alpha_s)$ and the condensate terms start at order $1 / (m_h^{pole})^4$~\cite{Boyd:1997kz}, we fit the lattice data for the double ratios adopting the following Ansatz
 \bea
    \label{eq:doubleratio_fit}
    \frac{R_j}{R_j^{PT}} & = & 1 + \left( \frac{\alpha_s(m_h^{pole})}{\pi} \right)^2 ~ \sum_{k=0}^3 A_k \left( \frac{m_c^{pole}}{m_h^{pole}} \right)^k \nonumber \\[2mm]
                                     & + & \left[ A_4 + A_4^s \frac{\alpha_s(m_h^{pole})}{\pi} \right] \left( \frac{m_c^{pole}}{m_h^{pole}} \right)^4 ~ , ~ \quad
\eea
which contains six parameters ($A_0$, $A_1$, $A_2$, $A_3$, $A_4$ and $A_4^s$) to be determined by a $\chi^2$-minimization procedure. 
We remind that, for sake of simplicity, we have dropped in the notation of all the parameters their dependence on the specific channel $j$.
The quality of the fits is illustrated in Figs.~\ref{fig:DRVLAL}-\ref{fig:DRVTAT}, where the values obtained at the physical $b$-quark point are shown as crosses.
The best value of $\chi^2 / ({\rm d.o.f.})$ turns out to be always well below $0.1$.
Within the uncertainties a nice agreement is found between the data as well as between the results of the fitting procedure corresponding to the two combinations $(r, -r)$ and $(r,r)$ of the Wilson $r$-parameters.
Therefore, in what follows we first average the lattice data over the two $r$-combinations and then perform the fit based on Eq.~(\ref{eq:doubleratio_fit}).

Using the double ratios the susceptibilities $\chi_j(m_b^{phys})$ can be expressed as
\be
    \label{eq:VL_mb}
    \chi_{0^+}(m_b^{phys}) = \chi_{0^+}(\lambda m_c^{phys}) \cdot \frac{\chi_{0^+}^{PT}(m_b^{phys})}{\chi_{0^+}^{PT}(\lambda m_c^{phys})} 
                                             \cdot  \prod_{n=3}^{11}  \frac{R_{0^+}(n)}{R_{0^+}^{PT}(n)} ~ , ~ 
\ee
and for $j = \{1^-, 0^-, 1^+ \}$
\be
   \label{eq:chi_mb}
    \chi_j(m_b^{phys}) = \chi_j(m_c^{phys}) \cdot \frac{\chi_j^{PT}(m_b^{phys})}{\chi_j^{PT}(m_c^{phys})} \cdot 
                                      \prod_{n=2}^{11}  \frac{R_j(n)}{R_j^{PT}(n)} ~ . ~
\ee
In Eqs.~(\ref{eq:VL_mb})-(\ref{eq:chi_mb}) the products over $n$ include the lattice data up to $n = 9$ and then the results of the fitting function (\ref{eq:doubleratio_fit}) for $n = 10$ and $n = 11$.
We remind that the case $j = 0^+$ is treated differently (i.e., $n \geq 3$ in Eq.~(\ref{eq:VL_mb})) because $\chi_{0^+}(m_c^{phys}) = \chi_{0^+}^{PT}(m_c^{phys}) = 0$ by charge conservation.

The results obtained for the ratios $\chi_{0^+}(m_b^{phys}) / \chi_{0^+}(\lambda m_c^{phys})$ and $\chi_j(m_b^{phys}) /$ $\chi_j(m_c^{phys})$ for $j = \{1^-, 0^-, 1^+ \}$ are shown in Table~\ref{tab:doubleratio} and compared with the NLO PT predictions based on Eq.~(\ref{eq:chi_PT}).
It can be seen that non-perturbative effects (as well as PT ones at order ${\cal{O}}(\alpha_s^2)$ and beyond) may affect the vector susceptibility ratios at the level of $\sim 10 - 20 \%$, while the axial ones at the level of few percent only.
\begin{table}[htb!]
\renewcommand{\arraystretch}{1.20}
\begin{center}
\begin{tabular}{||c||c|c||c|c||}
\hline
\multicolumn{1}{||c||}{} & \multicolumn{2}{|c||}{$1^{st}-4^{th}$ branches} & \multicolumn{2}{|c||}{$5^{th}-8^{th}$ branches} \\ \hline
susc.~ratio & PT & lattice & PT & lattice \\ \hline \hline
$0^+$ & ~~99.9~(0.8) & ~~89.5~(9.6) & ~101.3~(0.9) & ~~87.5~(5.4) \\ \hline
$1^-$ & 0.1931~~(7)  & 0.166~(10)    & 0.1949~~(7)  & 0.159~~(4) \\ \hline
$0^-$ & 0.7917~(10)  & 0.789~(35)    & 0.7922~(10)  & 0.760~(15) \\ \hline
$1^+$ & 0.2346~~(7)  & 0.245~(15)    & 0.2363~~(7)  & 0.239~~(7) \\ \hline \hline
\end{tabular}
\end{center}
\renewcommand{\arraystretch}{1.0}
\caption{\it \small Results for the susceptibility ratios $\chi_{0^+}(m_b^{phys}) / \chi_{0^+}(\lambda m_c^{phys})$ and $\chi_j(m_b^{phys}) /$ $\chi_j(m_c^{phys})$ for $j = \{1^-, 0^-, 1^+ \}$ based on Eqs.~(\ref{eq:VL_mb}) and (\ref{eq:chi_mb}), respectively, averaged over the $1^{st}-4^{th}$ and $5^{th}-8^{th}$ branches of our bootstrap analysis (see Appendix~\ref{sec:simulations}). The second and fourth columns represent the NLO PT ratios obtained using Eq.~(\ref{eq:chi_PT}). The third and fifth columns correspond to the  lattice data averaged over the two $r$-combinations and fitted using Eq.~(\ref{eq:doubleratio_fit}).}
\label{tab:doubleratio}
\end{table}

\begin{figure}[htb!]
\begin{center}
\includegraphics[scale=0.80]{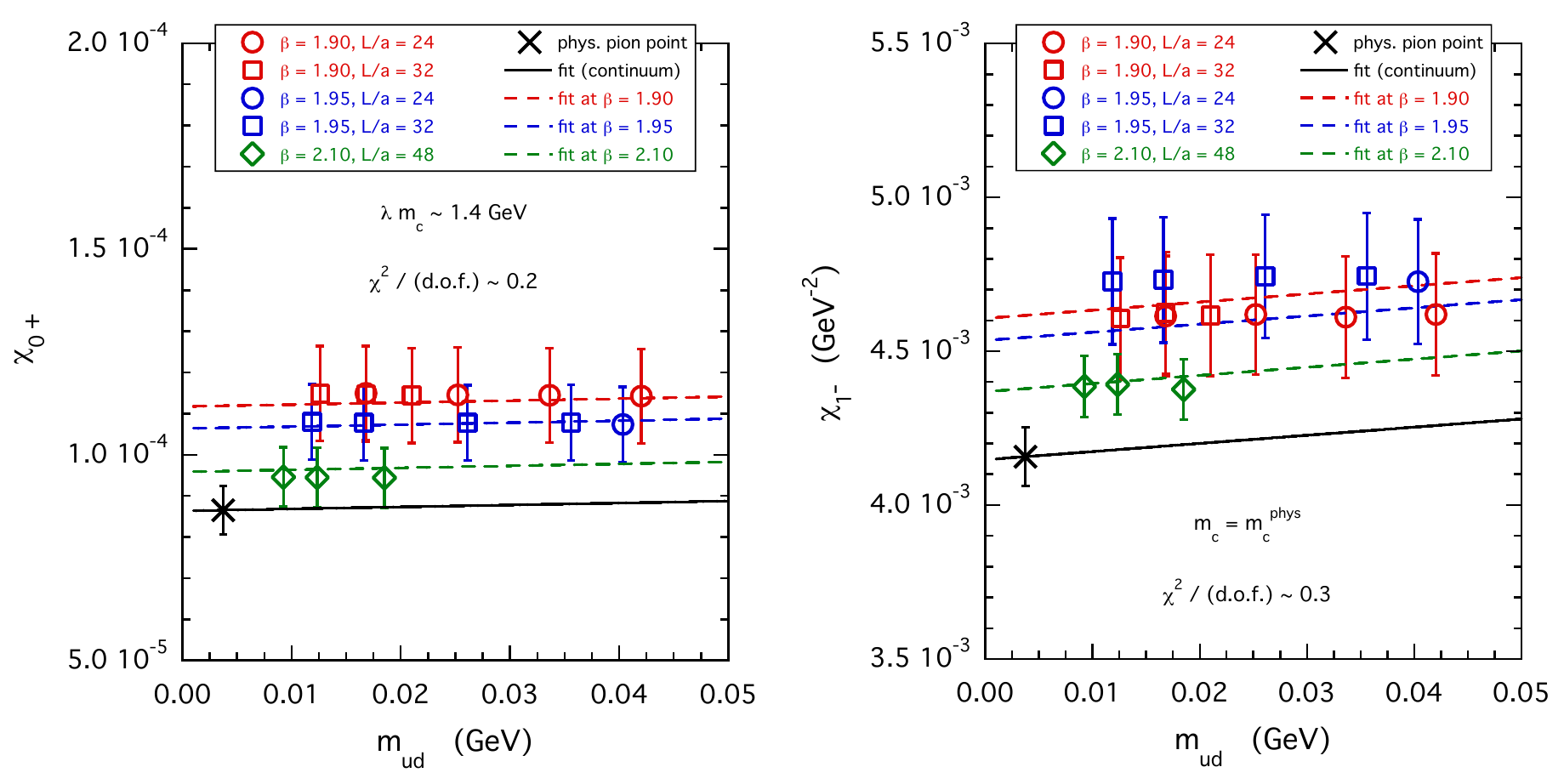}
\end{center}
\vspace{-1cm}
\caption{\it \small Light-quark mass dependence of the longitudinal susceptibilities $\chi_{0^+}(\lambda m_c^{phys})$ (left panel) and of the transverse one $\chi_{1^-}(m_c^{phys})$ (right panel), averaged over the two combinations $(r, -r)$ and $(r,r)$ of the Wilson $r$-parameters. The solid and dashed lines represent the results of a fitting function similar to the one adopted for the ETMC ratios~(\ref{eq:ratiofit}), including the same gaussian prior, evaluated respectively in the continuum limit and at each value of $\beta$. The crosses represent the values of the susceptibilities extrapolated to the physical pion point ($m_{ud} = m_{ud}^{phys}$) and to the continuum limit.}
\label{fig:VLVT_trig}
\end{figure}

In Eqs.~(\ref{eq:VL_mb})-(\ref{eq:chi_mb}) the ingredients that remain to be determined are the susceptibilities evaluated at the lightest heavy-quark mass (triggering point), namely $\chi_{0^+}(\lambda m_c^{phys})$ and $\chi_j(m_c^{phys})$ for $j = \{1^-, 0^-, 1^+ \}$.
The extrapolation to the physical pion point and to the continuum limit of the lattice data is performed using a fitting function similar to the one adopted for the ETMC ratios~(\ref{eq:ratiofit}) including the same gaussian prior.
The quality of the fits is illustrated in Fig.~\ref{fig:VLVT_trig} in the case of the longitudinal susceptibilities $\chi_{0^+}(\lambda m_c^{phys})$ and of the transverse one $\chi_{1^-}(m_c^{phys})$.

The values of the susceptibilities extrapolated to the physical pion point ($m_{ud} = m_{ud}^{phys}$) and to the continuum limit are collected in Table~\ref{tab:triggering} and compared with the corresponding PT predictions based on Eq.~(\ref{eq:chi_PT}).
It can be seen that non-perturbative effects (as well as PT ones at order ${\cal{O}}(\alpha_s^2)$ and beyond) affect significantly all the susceptibilities at the triggering point except the longitudinal axial one.
\begin{table}[htb!]
\renewcommand{\arraystretch}{1.20}
\begin{center}
\begin{tabular}{||c||c|c||c|c||}
\hline
\multicolumn{5}{||c||}{triggering~point} \\ \hline
\multicolumn{1}{||c||}{} & \multicolumn{2}{|c||}{$1^{st}-4^{th}$ branches} & \multicolumn{2}{|c||}{$5^{th}-8^{th}$ branches} \\ \cline{2-5}
$j$                               & $\chi_j^{PT}$ & $\chi_j$ & $\chi_j^{PT}$ & $\chi_j$ \\ \hline \hline
$0^+$                          & 4.50\,(3) $ \cdot 10^{-5}$ & 8.65\,(59) $ \cdot 10^{-5}$ &4.40\,(3) $ \cdot 10^{-5}$ & 8.46\,(30) $ \cdot 10^{-5}$ \\ \hline
$1^-$~(GeV$^{-2}$).   & 2.55\,(9) $ \cdot 10^{-3}$ & 4.16\,(15) $ \cdot 10^{-2}$ &2.64\,(9) $ \cdot 10^{-3}$ & 4.11\,(12) $ \cdot 10^{-2}$ \\ \hline
$0^-$                           & 3.14\,(1) $ \cdot 10^{-2}$ & 3.42\,~(9) $ \cdot 10^{-2}$ &3.14\,(1) $ \cdot 10^{-2}$ & 3.24\,~(6) $ \cdot 10^{-2}$ \\ \hline
$1^+$~(GeV$^{-2}$)   & 1.20\,(4) $ \cdot 10^{-3}$ & 1.95\,~(6) $ \cdot 10^{-3}$ &1.25\,(4) $ \cdot 10^{-3}$ & 1.93\,~(4) $ \cdot 10^{-3}$ \\ \hline \hline
\end{tabular}
\end{center}
\renewcommand{\arraystretch}{1.0}
\caption{\it \small Values of the longitudinal and transverse susceptibilities at the triggering point, namely $m_h = \lambda m_c^{phys}$ for $j = 0^+$ and $m_h = m_c^{phys}$ for $j = \{1^-, 0^-, 1^+ \}$, averaged over the $1^{st}-4^{th}$ and $5^{th}-8^{th}$ branches of our bootstrap analysis (see Appendix~\ref{sec:simulations}). The transverse susceptibilities are in physical units. The second and fourth columns represent the NLO PT susceptibilities obtained using Eq.~(\ref{eq:chi_PT}).}
\label{tab:triggering}
\end{table}

Using the results of Tables~\ref{tab:doubleratio} and \ref{tab:triggering} the longitudinal and transverse susceptibilities can be calculated at the physical $b$-quark point using Eqs.~(\ref{eq:VL_mb})-(\ref{eq:chi_mb}). 
Our findings are collected in Table~\ref{tab:chi_mb} and exhibit a remarkable difference with respect to the NLO PT predictions, based on Eq.~(\ref{eq:chi_PT}), except for the case of the longitudinal axial susceptibility.
\begin{table}[htb!]
\renewcommand{\arraystretch}{1.20}
\begin{center}
\begin{tabular}{||c||c|c||c|c||}
\hline
\multicolumn{1}{||c||}{} & \multicolumn{2}{|c||}{$1^{st}-4^{th}$ branches} & \multicolumn{2}{|c||}{$5^{th}-8^{th}$ branches} \\ \hline
$j$                               & $\chi_j^{PT}(m_b^{phys})$ & $\chi_j(m_b^{phys})$ & $\chi_j^{PT}(m_b^{phys})$ & $\chi_j(m_b^{phys})$ \\ \hline \hline
$0^+$                          & 4.49\,~(1) $ \cdot 10^{-3}$ & 7.75\,(73) $ \cdot 10^{-3}$ & 4.46\,~(1) $ \cdot 10^{-3}$ & 7.40\,(32) $ \cdot 10^{-3}$ \\ \hline
$1^-$~(GeV$^{-2}$)   & 4.92\,(15) $ \cdot 10^{-4}$ & 6.90\,(47) $ \cdot 10^{-4}$ & 5.15\,(16) $ \cdot 10^{-4}$ & 6.54\,(22) $ \cdot 10^{-4}$ \\ \hline
$0^-$                          & 2.48\,~(1) $ \cdot 10^{-2}$ & 2.70\,(14) $ \cdot 10^{-2}$ & 2.49\,~(1) $ \cdot 10^{-2}$ & 2.46\,~(9) $ \cdot 10^{-2}$ \\ \hline
$1^+$~(GeV$^{-2}$).  & 2.82\,~(9) $ \cdot 10^{-4}$ & 4.77\,(36) $ \cdot 10^{-4}$ & 2.95\,~(9) $ \cdot 10^{-4}$ & 4.60\,(19) $ \cdot 10^{-4}$ \\ \hline \hline
\end{tabular}
\end{center}
\renewcommand{\arraystretch}{1.0}
\caption{\it \small Values of the longitudinal and transverse susceptibilities at the physical $b$-quark point, given by Eqs.~(\ref{eq:VL_mb}-\ref{eq:chi_mb}), averaged over the $1^{st}-4^{th}$ and $5^{th}-8^{th}$ branches of our bootstrap analysis (see Appendix~\ref{sec:simulations}). The transverse susceptibilities are in physical units. The second and fourth columns represent the NLO PT susceptibilities obtained using Eq.~(\ref{eq:chi_PT}).}
\label{tab:chi_mb}
\end{table}

After averaging over all the eight branches of our bootstrap analysis (see Eq.~(\ref{eq:combineresults}) of Appendix~\ref{sec:simulations}) our final results are
\bea
     \label{eq:chiVL_final}
     \chi_{0^+}(m_b^{phys}) & = & 7.58 ~ (59) \cdot 10^{-3} ~ , ~ \\[2mm]
     \label{eq:chiVT_final}
     \chi_{1^-}(m_b^{phys}) & = & 6.72 ~ (41) \cdot 10^{-4} ~ \mbox{GeV}^{-2} ~ , ~ \\[2mm]
     \label{eq:chiAL_final}
     \chi_{0^-}(m_b^{phys}) & = & 2.58 ~ (17) \cdot 10^{-2} ~ , ~ \\[2mm]
     \label{eq:chiAT_final}
     \chi_{1^+}(m_b^{phys}) & = & 4.69 ~ (30) \cdot 10^{-4} ~ \mbox{GeV}^{-2} ~ . ~
\eea

In Refs.~\cite{Bigi:2016mdz,Bigi:2017njr,Bigi:2017jbd} the susceptibilities $\chi_j(m_b^{phys})$ have been estimated using PT at NNLO, using physical values of the $c$- and $b$-quark masses lower than our ETMC values, namely a $c$-quark mass lower by $\simeq 7.3 \%$ and a $b$-quark mass lower by $\simeq 2.4 \%$.
The values obtained for the susceptibilities were: $\chi_{0^+}(m_b^{phys}) = 6.204~(81) \cdot 10^{-3}$,  $\chi_{1^-}(m_b^{phys}) = 6.486~(48) \cdot 10^{-4}$ GeV$^{-2}$, $\chi_{0^-}(m_b^{phys}) = 2.41 \cdot 10^{-2}$ and $\chi_{1^+}(m_b^{phys}) = 3.894 \cdot 10^{-4}$ GeV$^{-2}$.
While the transverse vector and longitudinal axial susceptibilities are only $\simeq 4 \%$ and $\simeq 7 \%$ lower than our findings (\ref{eq:chiVT_final}) and (\ref{eq:chiAL_final}), the longitudinal vector and the transverse axial susceptibilities are $\simeq 20 \%$ lower than our results (\ref{eq:chiVL_final}) and (\ref{eq:chiAT_final}), respectively. 
In all cases the differences are within $\sim 2.5$ standard deviations.

\subsection{Alternative analysis}
\label{sec:alternative}

The results obtained in the previous Section suggests that the heavy-quark mass dependence of the ratios $R_j$ is mainly dictated by the corresponding NLO PT predictions $R_j^{PT}$.
We want to check further this point and we perform a direct fit of the ratios $R_j$ using the following Ansatz
 \be
    \label{eq:ratio_fit}
   R_j = 1 + \sum_{k = 1}^3 \left[ A_k + A_k^s  \frac{\alpha_s(m_h)}{\pi} \right] \left( \frac{m_c}{m_h} \right)^k ~ , ~
\ee
which contains six parameters ($A_1$, $A_1^s$, $A_2$, $A_2^s$, $A_3$ and $A_3^s$) to be determined by a $\chi^2$-minimization procedure.
Note that, at variance with Eq.~(\ref{eq:doubleratio_fit}) of the previous Section, in Eq.~(\ref{eq:ratio_fit}) the running mass $m_h$ is adopted.
Since the condensate terms start at order $1 / m_h^4$~\cite{Boyd:1997kz}, the fitting function (\ref{eq:ratio_fit}) is expected to represent $R_j^{PT}$ up to order $1 / m_h^3$.
The quality of the fits is as good as the one found for the double ratios $R_j / R_j ^{PT}$ and the best value of $\chi^2 / ({\rm d.o.f.})$ does not exceed $\sim 0.1$.

Using the ratios $R_j$ the susceptibilities $\chi_j(m_b^{phys})$ at the physical $b$-quark point can be expressed as
\bea
    \label{eq:VL_mb_alt}
    \chi_{0^+}(m_b^{phys}) & = & \chi_{0^+}(\lambda m_c^{phys}) \cdot  \prod_{n=3}^{11} R_{0^+}(n) ~ , ~ \\[2mm]
    \chi_{0^-}(m_b^{phys}) & = & \chi_{0^-}(m_c^{phys}) \cdot \prod_{n=2}^{11} R_{0^-}(n) 
    \label{eq:AL_mb_alt}
\eea
and for $j = \{1^-, 1^+ \}$
\be
   \label{eq:chiT_mb_alt}
    \chi_j(m_b^{phys}) = \left[ \frac{m_c^{pole}(m_c^{phys})}{m_b^{pole}(m_b^{phys})} \right]^2 \chi_j(m_c^{phys}) \cdot \prod_{n=2}^{11} R_j(n) ~ . ~
\ee
After averaging over all the eight branches of our bootstrap analysis we quote the final values obtained for the susceptibilities $\chi_j(m_b^{phys})$, namely
 \bea
     \label{eq:chiVL_final_alt}
     \chi_{0^+}(m_b^{phys}) & = & 7.52 ~ (63) \cdot 10^{-3} ~ , ~ \\[2mm]
     \label{eq:chiVT_final_alt}
     \chi_{1^-}(m_b^{phys}) & = & 6.76 ~ (40) \cdot 10^{-4} ~ \mbox{GeV}^{-2} ~ , ~ \\[2mm]
     \label{eq:chiAL_final_alt}
     \chi_{0^-}(m_b^{phys}) & = & 2.59 ~ (18) \cdot 10^{-2} ~ , ~ \\[2mm]
     \label{eq:chiAT_final_alt}
     \chi_{1^+}(m_b^{phys}) & = & 4.68 ~ (30) \cdot 10^{-4} ~ \mbox{GeV}^{-2} ~ , ~
\eea
which nicely agrees (in a reassuring way) with the findings (\ref{eq:chiVL_final})-(\ref{eq:chiAT_final}).

\begin{figure}[htb!]
\begin{center}
\includegraphics[scale=0.80]{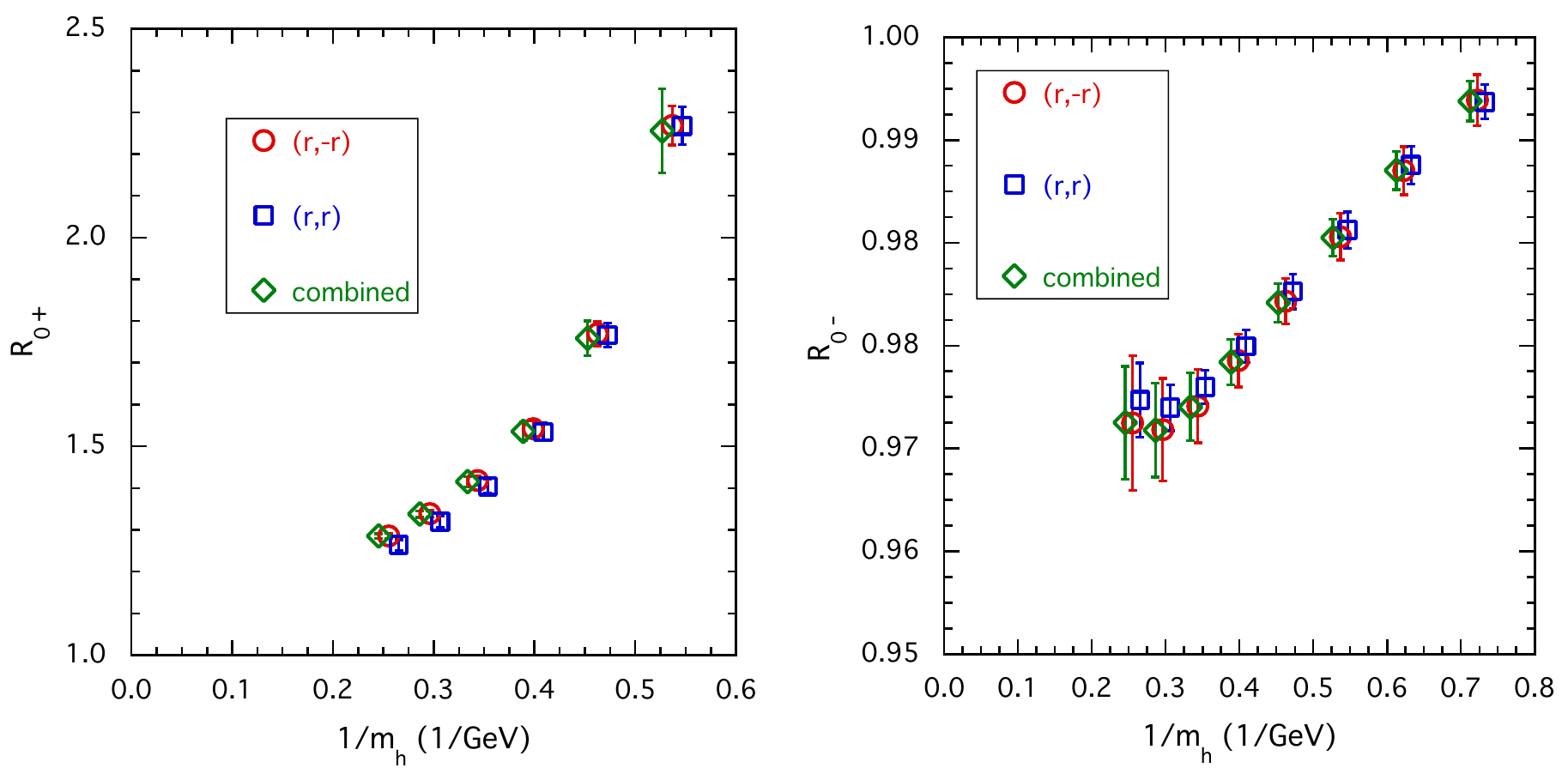}
\end{center}
\vspace{-1cm}
\caption{\it \small The longitudinal susceptibility ratios $R_{0^+}$ (left panel) and $R_{0^-}$ (right panel), after extrapolation to the physical pion point and to the continuum limit based on the combined fit~(\ref{eq:combinedfit}) of the data corresponding to the two combinations $(r, -r)$ and $(r,r)$ of the Wilson $r$-parameters, versus the inverse heavy-quark mass $1 / m_h$. The dots and squares correspond to the longitudinal susceptibility ratios obtained in Section~\ref{sec:ETMC_ratio} using the fitting function (\ref{eq:ratiofit}) and adopting the value $D_{prior} = 20$ (see Fig.~\ref{fig:RVLAL}). Note that the locations of the results are slightly moved away for a better reading.}
\label{fig:RVLAL_combined}
%\end{figure}
%\begin{figure}[htb!]
\begin{center}
\includegraphics[scale=0.80]{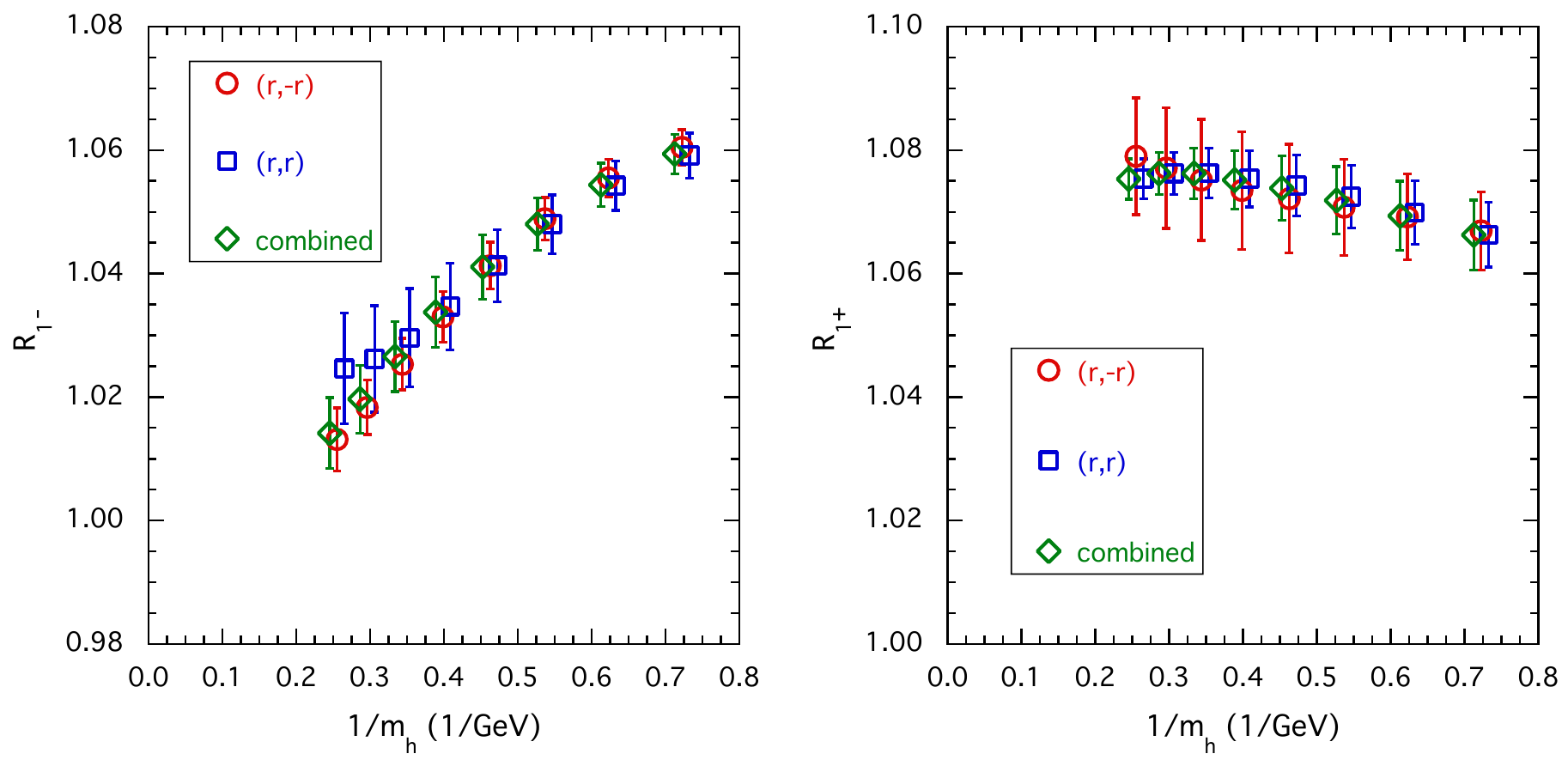}
\end{center}
\vspace{-1cm}
\caption{\it \small The same as in Fig.~\ref{fig:RVLAL_combined}, but in the case of the transverse susceptibility ratios $R_{1^-}$ (left panel) and $R_{1^+}$ (right panel).}
\label{fig:RVTAT_combined}
\end{figure}

Another check is dictated by the fact that the susceptibilities corresponding to the two combinations $(r, -r)$ and $(r,r)$ of the Wilson $r$-parameters should differ only by discretization effects starting at order ${\cal{O}}(a^2)$.
This suggests to perform a combined extrapolation to the continuum limit (and to the physical pion point) of both combinations.
In terms of the ratios  $R_j^{(r, \pm r)}(n; a^2, m_{ud})$, defined as in Eq.~(\ref{eq:ETMC_ratios}), one has 
\bea
    \label{eq:combinedfit}
    R_j^{(r, \pm r)}(n; a^2, m_{ud}) & = & R_j(n) \left[ 1 + A_1 \left( m_{ud} - m_{ud}^{phys} \right) \right. \nonumber \\[2mm]
                                                      & + & \left. D_1^{(r, \pm r)} ~ \frac{a^2}{r_0^2} + D_2^{(r, \pm r)} ~ \frac{a^4}{r_0^4} \right] ~ , ~
\eea
without including any prior at variance with what done in Eq.~(\ref{eq:ratiofit}).
The quality of the fitting procedure is always very good for all channels ($\chi^2 / ({\rm d.o.f.}) < 0.3$) and the results obtained for $R_j(n)$ are shown in Figs.~\ref{fig:RVLAL_combined}-\ref{fig:RVTAT_combined} as green diamonds, where they are compared with those previously determined using Eq.~(\ref{eq:ratiofit}) for the two $r$-combinations separately.
The agreement is remarkably good and quite reassuring about the control of the extrapolation of our ratios to the continuum limit.

\section{Subtraction of the ground-state contribution}
\label{sec:grs}

The susceptibilities $ \chi_j(m_b^{phys})$ for $j = \{ 0^+, 1^-, 0^-, 1^+  \}$ represent upper limits to the dispersive bounds on the form factors relevant in the semileptonic $B \to D(D^*) \ell \nu_\ell$ decays.
Such limits can be improved by removing the contributions of the bound states from the calculated Euclidean correlators $C_j(t)$ (see Eqs.~(\ref{eq:CVL12}-\ref{eq:CP12}) for $j = \{ 0^+, 1^-, 0^-, 1^+, S, P \}$).
In this Section we describe the subtraction of the ground-state contribution.

As well known, at large time distances one has
\be
    \label{eq:larget} 
     C_j(t)_{ ~ \overrightarrow{t  \gg a, ~ (T - t) \gg a} ~ } \frac{\mathcal{Z}_j}{2M_j} \left[ e^{ - M_j  t} + e^{ - M_j (T - t)} \right] ~ ,  
\ee
where $M_j$ is the mass of ground-state meson $H_{12}^j$ and $\mathcal{Z}_j$ is the matrix element $ \mathcal{Z}_j \equiv | \langle H_{12}^j | \overline{q}_1 \Gamma_j q_2 | 0 \rangle|^2$ with $\Gamma_j = \{ \gamma_0, \vec{\gamma}, \gamma_0 \gamma_5, \vec{ \gamma} \gamma_5, \mathbb{1} , \gamma_5 \}$ for $j = \{ 0^+, 1^-, 0^-, 1^+, S, P \}$. 

\begin{figure}[htb!]
\begin{center}
\includegraphics[scale=0.80]{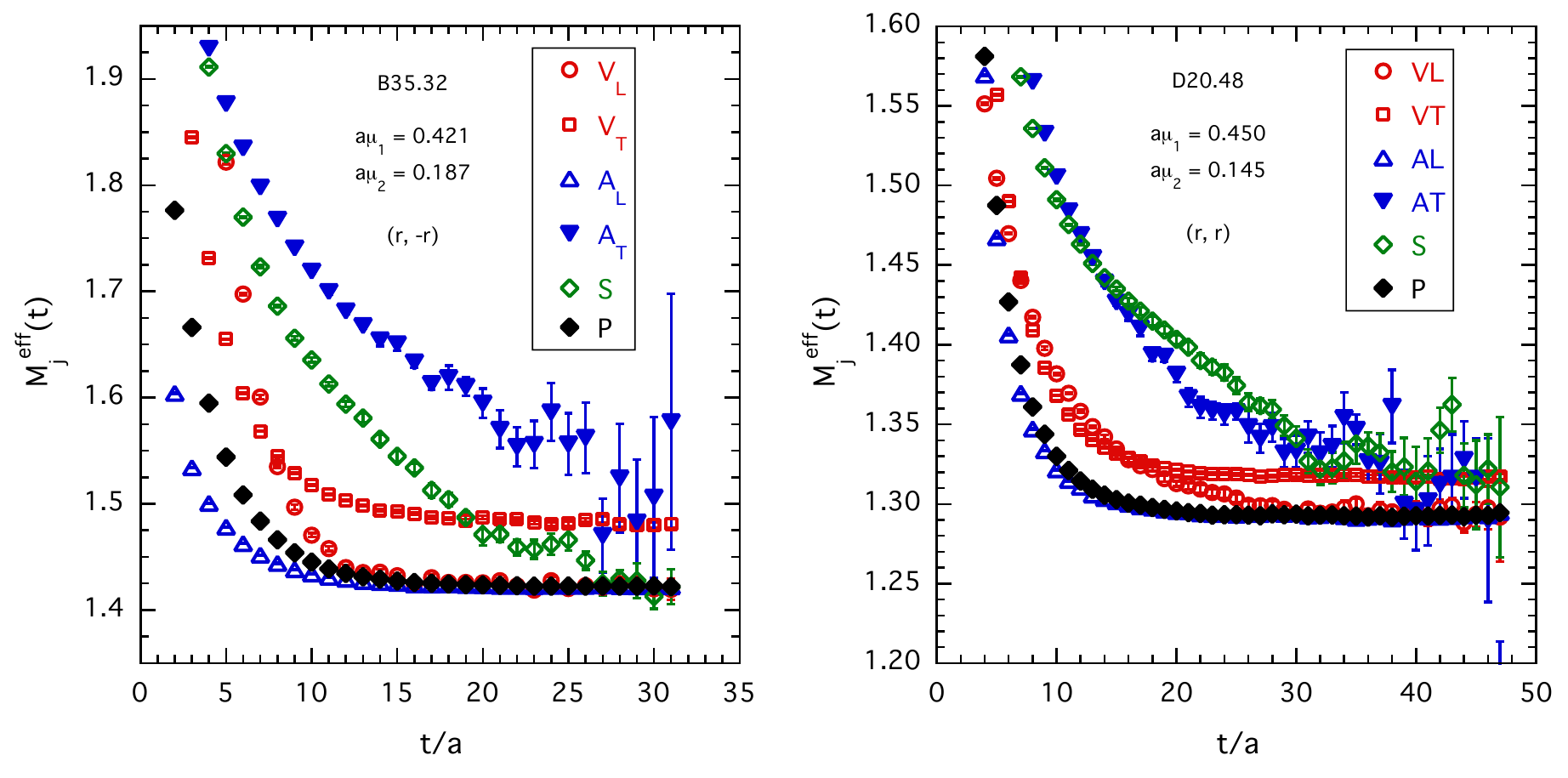}
\end{center}
\vspace{-1cm}
\caption{\it \small The temporal behavior of the effective mass (\ref{eq:Meff}) for the six correlators $C_j(t)$ with $j = \{ 0^+, 1^-, 0^-, 1^+ , S, P \}$ in the case of the ensembles B35.32 (left panel) and D20.48 (right panel). The (bare) quark masses and the combinations of the Wilson $r$-parameters are specified in the insets and they roughly correspond to the case $m_2 \approx m_c^{phys}$ with $m_1 \approx 2 m_c^{phys}$ (left panel) and $m_1 \approx 3 m_c^{phys}$ (right panel).}
\label{fig:plateaux}
\end{figure}

Thus, the ground-state mass $M_j$ and the matrix element $\mathcal{Z}_j$ can be extracted from the exponential fit given in the r.h.s.~of Eq.~(\ref{eq:larget}) performed in the temporal region $ t = [t_{min}, t_{max}]$, where the effective mass $M_j^{eff}(t)$
\be
    \label{eq:Meff}
    M_j^{eff}(t) \equiv \mbox{log}\left( \frac{C_j(t-1)}{C_j(t)} \right)
\ee
exhibits a plateau.
The quality of the plateaux for the effective masses $M_j^{eff}(t)$ is shown in Fig.~\ref{fig:plateaux} in two illustrative cases.
The quality of the plateaux is good for $j = 0^+, 1^-, 0^-$ and $j = P$, while it is very poor in the cases $j = 1^+$ and $j = S$.
The latter case is likely to be plagued by the effects of parity breaking (mixing with $j = P$) present in our lattice formulation.
The same is expected to occur for $j = 0^+$ in spite of the good plateaux observed.
Therefore, in what follows we limit ourselves to the analysis of the transverse vector $C_{1^-}(t)$ and longitudinal axial $C_{0^-}(t)$ correlators. 
The temporal regions chosen for performing the exponential fit given in the r.h.s.~of Eq.~(\ref{eq:larget}) are shown in Table~\ref{tab:plateaux} for the various ETMC ensembles.
\begin{table}[htb!]
\begin{center}	
\begin{tabular}{||c|c||c||}
\hline
$\beta$ & $V / a^4$ & $[t_{\rm min} / a, \, t_{\rm max} / a]$ \\
\hline \hline
1.90 & $24^3 \times 48$ & [18, 22] \\ \cline{2-3}
        & $32^3 \times 64$ & [18, 28] \\ \hline
1.95 & $24^3 \times 48$ & [19, 22] \\ \cline{2-3}
        & $32^3 \times 64$ & [19, 28] \\ \hline
2.10 & $48^3 \times 96$ & [24, 38] \\ \hline
\end{tabular}
\end{center}
\caption{\it Temporal regions chosen for performing the exponential fit given in the r.h.s.~of Eq.~(\ref{eq:larget}) for the various ETMC ensembles.}
\label{tab:plateaux}
\end{table}

Since we make use of the WIs for the longitudinal axial-vector susceptibility (see Eq.~(\ref{eq:chiAL_WI0})), the ground-state contributions $\chi_{1^-}^{(gs)}(m_1, m_2)$ and $\chi_{0^-}^{(gs)}(m_1, m_2)$ are explicitly given by
\bea
      \label{eq:chiVT_grs}
      \chi_{1^-}^{(gs)}(m_1, m_2) & = & \frac{f_V^2}{M_V^4} ~ , ~ \\[2mm]
      \label{eq:chiAL_grs}
      \chi_{0^-}^{(gs)}(m_1, m_2) & = & \frac{f_P^2}{M_P^2} ~ , ~    
\eea
where $M_{V(P)}$ is the ground-state mass and $f_{V(P)}$ is the (leptonic) decay constant of a vector (pseudoscalar) meson made by two valence quarks of (renormalized) masses $m_1$ and $m_2$, respectively\footnote{The relation between $f_{V(P)}$ and the matrix element $\mathcal{Z}_{V(P)}$ appearing in Eq.~(\ref{eq:larget}) is $f_V^2 = \mathcal{Z}_V / M_V^2$ and $f_P^2 = (\mathcal{Z}_P / M_P^2) [ (am_1 + am_2) / (\mbox{sinh}(aM_P))]^2$.}. 

\begin{figure}[htb!]
\begin{center}
\includegraphics[scale=0.80]{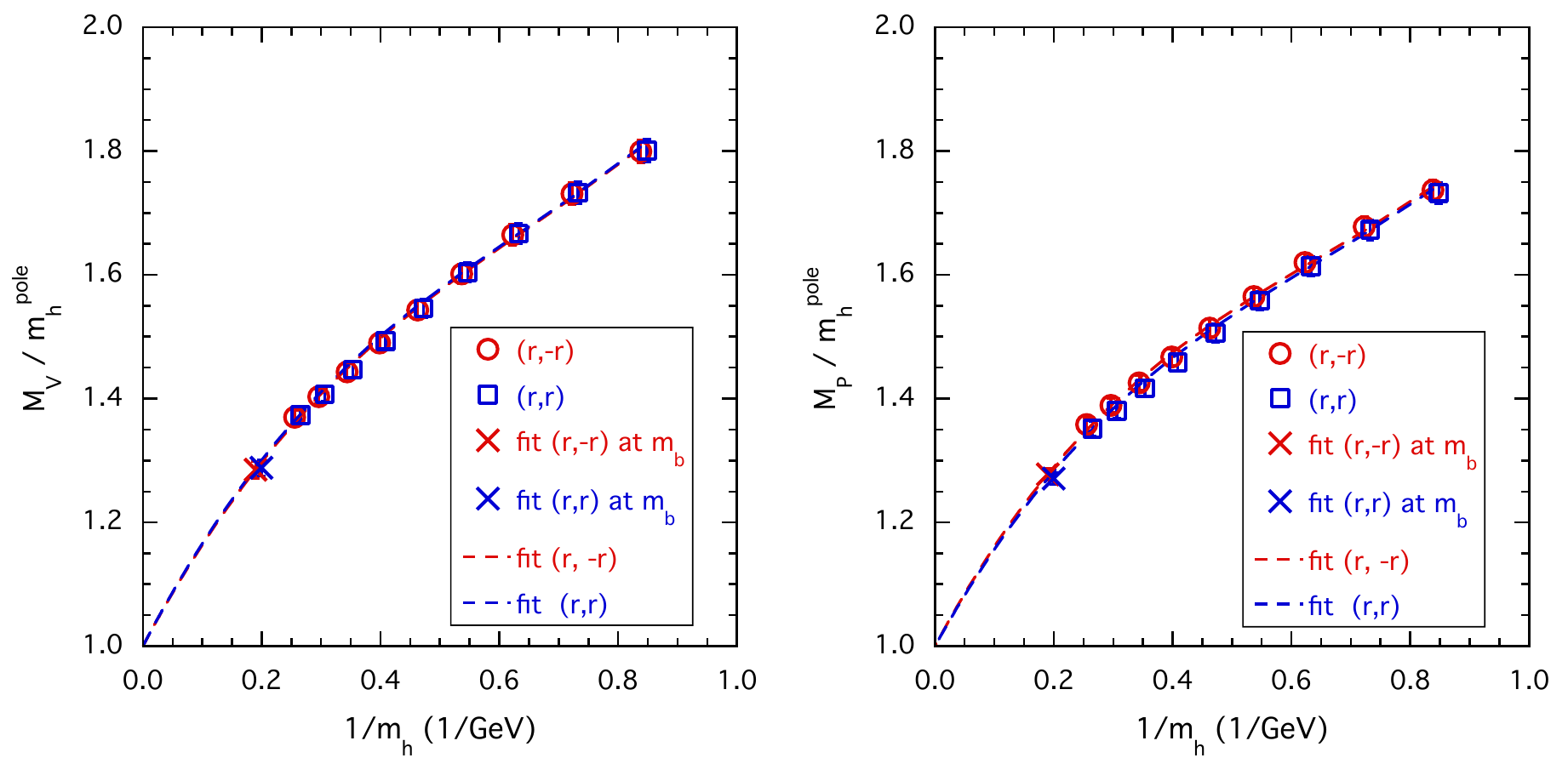}
\end{center}
\vspace{-1cm}
\caption{\it \small Vector (left panel) and pseudoscalar (right panel) ground-state masses, divided by the pole quark mass $m_h^{pole}$, versus the inverse heavy-quark mass $1 / m_h$. The dashed lines represent the results obtained with the fitting function (\ref{eq:massratios_fit}), while the crosses are the fit values at the physical $b$-quark point, indicated by vertical dotted lines. We remind that the locations of the results for the combination $(r, r)$ are slightly moved to the right for a better reading.}
\label{fig:MVTAL}
\end{figure}

We start by interpolating the ground-state masses $M_{j = V, P}(m_1, m_2)$ at $m_1 = m_h$ with $m_h$ given by the series of values~(\ref{eq:mh_n}) and at $m_2 = m_c^{phys}$.
Then, we extrapolate the masses to the physical pion point and to the continuum limit using the fitting function
\be
    \label{eq:mass_fit}
    M_j(m_h; a^2, m_{ud}) = M_j(m_h) \left[ 1 + A_1^M \left( m_{ud} - m_{ud}^{phys} \right) + D_1^M ~ \frac{a^2}{r_0^2} + D_2^M ~ \frac{a^4}{r_0^4} \right] ~ 
\ee
and including a (gaussian) prior, $D_{prior}^M$, on the two parameters $D_1^M$ and $D_2^M$, as in the case of the ETMC ratios of Section~\ref{sec:ETMC_ratio}.
The value $D_{prior}^M = 5$ turns out to be sufficient to guarantee that the extrapolated values $M_j(m_h)$ corresponding to the two combinations $(r, -r)$ and $(r,r)$ of the Wilson $r$-parameters coincide within the uncertainties.
The results for $M_V(m_h)$ and $M_P(m_h)$, divided by the pole quark mass $m_h^{pole}(m_h)$, are shown in Fig.~\ref{fig:MVTAL}. 
Since in the heavy-quark limit $m_h^{pole} \to \infty$ the ratios $M_j(m_h) / m_h^{pole}(m_h)$ goes toward unity, we adopt the simple fitting function
\be
    \label{eq:massratios_fit}
    \frac{M_j(m_h)}{m_h^{pole}(m_h)} = 1 + A_1 \frac{m_c^{pole}}{m_h^{pole}} + \sum_{k = 2}^3 \left[ A_k + A_k^s  \frac{\alpha_s(m_h^{pole})}{\pi} \right] 
                                                               \left( \frac{m_c^{pole}}{m_h^{pole}} \right)^k ~ . ~
\ee
The quality of the above fit is quite good for both $j = V$ and $j = P$.
The corresponding results are shown by the dashed lines in Fig.~\ref{fig:MVTAL}, where the values obtained at the physical $b$-quark point are also shown as crosses.
Our findings at the the physical $c$- and $b$-quark points are collected in Table~\ref{tab:masses} and compared with the experimental results from the PDG~\cite{PDG} and the vector meson mass $M_{B_c^*}$ determined by the HPQCD Collaboration in Ref.~\cite{Dowdall:2012ab}.
\begin{table}[htb!]
\renewcommand{\arraystretch}{1.20}
\begin{center}
\begin{tabular}{||c||c|c||c||c||}
\hline
                                         & $1^{st}-4^{th}$ & $5^{th}-8^{th}$ & $1^{st}-8^{th}$ & exp. / lat. \\ 
                                         & branches & branches & branches & \\ \hline \hline
$M_{\eta_c}$~(GeV)        & 3.02\,(5)   & 3.02\,(5)  & 3.02\,(5)   & 2.9839\,(5)~\cite{PDG} \\ \hline
$M_{J/\psi}$~(GeV)         & 3.13\,(5)   & 3.13\,(4)   & 3.13\,(5)   & 3.096900\,(6)~\cite{PDG} \\ \hline \hline
$M_{B_c}$~(GeV)           & 6.36\,(15) & 6.48\,(14) & 6.42\,(16) & 6.2749\,(8)~\cite{PDG} \\ \hline
$M_{B_c^*}$~(GeV)        & 6.33\,(16) & 6.56\,(14) & 6.45\,(19) & 6.332\,(9)~\cite{Dowdall:2012ab} \\ \hline \hline
$M_{B_c} / M_{\eta_c}$   & 2.10\,(2)  & 2.15\,(2)   & 2.12\,(3)    & 2.10292\,(35)~\cite{PDG} \\ \hline
$M_{B_c^*} / M_{J/\psi}$ & 2.02\,(4)  & 2.10\,(2)   & 2.06\,(5)    & 2.0437\,(29)~\cite{PDG,Dowdall:2012ab} \\ \hline
\end{tabular}
\end{center}
\renewcommand{\arraystretch}{1.0}
\caption{\it \small Values of the vector and pseudoscalar ground-state masses at the physical $c$- and $b$-quark points, averaged over the $1^{st}-4^{th}$ and $5^{th}-8^{th}$ branches of our bootstrap analysis (see Appendix~\ref{sec:simulations}). The fourth column represents the average over all the eight branches of our bootstrap analysis. The last column contains the experimental results from the PDG~\cite{PDG} and the vector meson mass $M_{B_c^*}$ determined by the HPQCD Collaboration in Ref.~\cite{Dowdall:2012ab}.}
\label{tab:masses}
\end{table}
A reasonable agreement within $\approx 1$ standard deviation can be observed for all the vector and pseudoscalar meson masses.
Since our physical $c$- and $b$-quark points are correlated, we can determine more accurately the pseudoscalar $M_{B_c} / M_{\eta_c}$ and the vector $M_{B_c^*} / M_{J/\psi}$ ratios, as shown in the last two rows of Table~\ref{tab:masses}.

\begin{figure}[htb!]
\begin{center}
\includegraphics[scale=0.80]{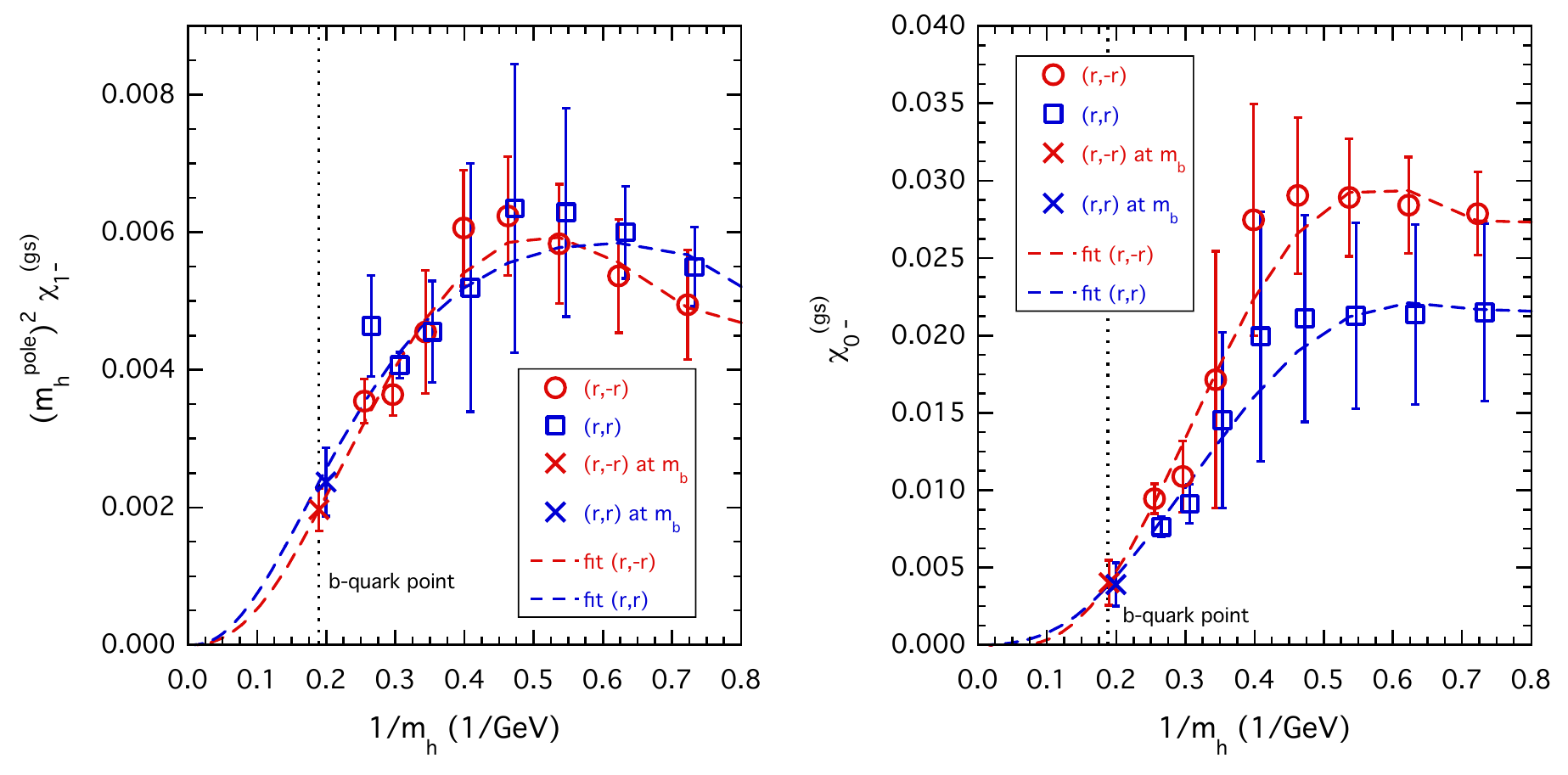}
\end{center}
\vspace{-1cm}
\caption{\it \small Ground-state susceptibilities $(m_h^{pole})^2 \chi_{1^-}^{(gs)}$ (left panel) and $\chi_{0^-}^{(gs)}$ (right panel) versus the inverse heavy-quark mass $1 / m_h$. The dashed lines represent the results obtained with the fitting function (\ref{eq:chigs_fit}), while the crosses are the fit values at the physical $b$-quark point, indicated by vertical dotted lines.}
\label{fig:VTAL_grs}
\end{figure}

As for the susceptibilities $\chi_j^{(gs)}(m_1, m_2)$, we first interpolate them at $m_1 = m_h$ with $m_h$ given by the series of values~(\ref{eq:mh_n}) and at $m_2 = m_c^{phys}$.
Then, for each combination of heavy-quark and charm masses we extrapolate the susceptibilities to the physical pion point and to the continuum limit using the fitting function
\bea
    \label{eq:chigs_hc}
    \chi_j^{(gs)}(m_h, m_c^{phys}; a^2, m_{ud}) & = & \chi_j^{(gs)}(m_h, m_c^{phys}) \left[ 1 + A_1^{(gs)} \left( m_{ud} - m_{ud}^{phys} \right) \right. \nonumber \\[2mm]
                                                                           & + & \left. D_1^{(gs)} ~ \frac{a^2}{r_0^2} + D_2^{(gs)} ~ \frac{a^4}{r_0^4} \right] ~ 
\eea
and including a (gaussian) prior, $D_{prior}^{(gs)}$, on the two parameters $D_1^{(gs)}$ and $D_2^{(gs)}$.
The value $D_{prior}^{(gs)} = 100$ turns out to be necessary to guarantee that the extrapolated values $\chi_j^{(gs)}(m_h, m_c^{phys})$ corresponding to the two combinations $(r, -r)$ and $(r,r)$ of the Wilson $r$-parameters coincide within the uncertainties.
The results obtained for $\chi_j^{(gs)}(m_h, m_c^{phys})$ are shown in Fig.~\ref{fig:VTAL_grs}, where we have multiplied the susceptibility $\chi_{1^-}^{(gs)}$ by $[m_h^{pole}(m_h)]^2$ in order to deal with dimensionless quantities.

For the extrapolation to the physical $b$-quark point we assume the validity of the scaling law $f_j \sqrt{M_j} = \mbox{const.}$ in the heavy-quark limit $m_h \to \infty$, which implies that both $(m_h^{pole})^2 \chi_{1^-}^{(gs)}$ and $\chi_{0^-}^{(gs)}$ should be at least of order ${\cal{O}}(1 / m_h^3)$.
Therefore, we adopt the following Ansatz
\be
    \label{eq:chigs_fit}
    \{ (m_h^{pole})^2 \chi_{1^-}^{(gs)}, \chi_{0^-}^{(gs)} \} \to \sum_{k = 3}^5 \left[ A_k + A_k^s  \frac{\alpha_s(m_h^{pole})}{\pi} \right] \left( \frac{m_c^{pole}}{m_h^{pole}} \right)^k ~ . ~
\ee
The quality of the fits is illustrated in Fig.~\ref{fig:VTAL_grs}, where the values obtained at the physical $b$-quark point are shown as crosses.
The best value of $\chi^2 / ({\rm d.o.f.})$ turns out to be in the range $0.3 - 1.1$.
Within the uncertainties a nice agreement is found between the results of the fitting procedure corresponding to the two combinations $(r, -r)$ and $(r,r)$ of the Wilson $r$-parameters.
Therefore, we average the lattice data over the two $r$-combinations and perform again the fit based on Eq.~(\ref{eq:chigs_fit}).
The results corresponding to the physical $b$-quark point are collected in Table~\ref{tab:chigs}.
\begin{table}[htb!]
\renewcommand{\arraystretch}{1.20}
\begin{center}
\begin{tabular}{||c||c|c||c||}
\hline
                                  & $1^{st}-4^{th}$ & $5^{th}-8^{th}$ & $1^{st}-8^{th}$ \\
                                  & branches & branches & branches \\ \hline \hline
$\chi_{1^-}^{(gs)} \cdot 10^4$~(GeV$^{-2}$) & 0.897\,(162) & 0.865\,(124) & 0.881\,(145) \\ \hline
$\chi_{0^-}^{(gs)} \cdot 10^2$                        & 0.400\,~(97) & 0.377\,~(86) & 0.389\,~(93) \\ \hline
\end{tabular}
\end{center}
\renewcommand{\arraystretch}{1.0}
\caption{\it \small Values of the ground-state susceptibilities at the physical $b$-quark point, averaged over the $1^{st}-4^{th}$ and $5^{th}-8^{th}$ branches of our bootstrap analysis (see Appendix~\ref{sec:simulations}). The transverse susceptibilities are in physical units. The last column represents the average over all the eight branches of our bootstrap analysis according to Eq.~(\ref{eq:combineresults}).}
\label{tab:chigs}
\end{table}

Our findings can be compared with those corresponding to the direct use of Eqs.~(\ref{eq:chiVT_grs})-(\ref{eq:chiAL_grs}) adopting the experimental result for $M_P = M_{B_c} = 6.2749\,(8)$ GeV from the PDG~\cite{PDG} and the results of the HPQCD Collaboration~\cite{Dowdall:2012ab,McNeile:2012qf,Colquhoun:2015oha,McLean:2019sds} for the mass $M_V = M_{B_c^*}$ as well as for the decay constants $f_V = f_{B_c^*}$ and $f_P = f_{B_c}$, namely: $M_{B_c^*} = 6.332\,(9)$ GeV, $f_{B_c^*} = 0.413\,(11)$ GeV and $f_{B_c} = 0.418\,(5)$ GeV, obtaining $\chi_{1^-}^{(gs)} = 1. 06\,(6) \cdot 10^{-4}$ GeV$^{-2}$ and $\chi_{0^-}^{(gs)} = 0.444\,(11) \cdot 10^{-2}$.
Thus, we observe an interesting (and reassuring) agreement within one standard deviation\footnote{In Refs.~\cite{Bigi:2016mdz,Bigi:2017njr} the values $\chi_{1^-}^{(gs)} = 1. 04 \cdot 10^{-4}$ GeV$^{-2}$ and $\chi_{0^-}^{(gs)} = 0.45 \cdot 10^{-2}$ were adopted.}.

As argued in Ref.~\cite{Caprini:1997mu}, the ground-state contributions $\chi_{0^+}^{(gs)}$ and $\chi_{1^-}^{(gs)}$ are expected to be small and they can be conservatively ignored.
For the same reason we can ignore also the subtraction of the contributions coming from higher excited bound-states.
Thus, after the subtraction of the ground-state contributions given in the last column of Table~\ref{tab:chigs} our non-perturbative lattice results for the susceptibilities relevant for the semileptonic $B \to D(D^*) \ell \nu_\ell$ decays are
\bea
     \label{eq:chiVL_BD}
     \chi_{0^+}(m_b^{phys}) & = & 7.58 ~ (59) \cdot 10^{-3} ~ , ~ \\[2mm]
     \label{eq:chiVT_BD}
     \chi_{1^-}^{sub}(m_b^{phys}) & = & 5.84 ~ (44) \cdot 10^{-4} ~ \mbox{GeV}^{-2} ~ , ~ \\[2mm]
     \label{eq:chiAL_BD}
     \chi_{0^-}^{sub}(m_b^{phys}) & = & 2.19 ~ (19) \cdot 10^{-2} ~ , ~ \\[2mm]
     \label{eq:chiAT_BD}
     \chi_{1^+}(m_b^{phys}) & = & 4.69 ~ (30) \cdot 10^{-4} ~ \mbox{GeV}^{-2} ~ . ~
\eea

\section{Conclusions}
\label{sec:conclusions}
In this work we have presented the first non-perturbative determination of the dispersive bounds that constrain the form factors entering the semileptonic $B \to D^{(*)} \ell \nu_\ell $ transitions due to unitarity and analyticity. 
The bounds are obtained by evaluating moments of suitable two-point correlation functions obtained on the lattice according to the dispersive method of Ref.~\cite{DiCarlo:2021dzg}.

By adopting the gauge ensembles produced by the Extended Twisted Mass Collaboration with $N_f = 2+1+1$ dynamical quarks at three values of the lattice spacing ($a \simeq 0.062, 0.082, 0.089$ fm) and with pion masses in the range $\simeq 210 - 450$ MeV, we have evaluated the longitudinal and transverse susceptibilities of the vector and axial-vector polarization functions at the physical pion point and in the continuum and infinite volume limits. 

The ETMC ratio method of Ref.~\cite{Blossier:2009hg} has been adopted to reach the physical $b$-quark mass $m_b^{phys}$, and the one-particle contributions due to $B_c^*$- and $B_c$-mesons are evaluated and subtracted to obtain improved bounds in the case of the vector and pseudoscalar channels.
At zero momentum transfer for the scalar, vector, pseudoscalar and axial susceptibilities we obtain the final results (\ref{eq:chiVL_BD})-(\ref{eq:chiAT_BD}),
% \bea
%    \label{eq:chi0+_final}
%     \chi_{0^+}(m_b^{phys}) & = & 7.58\,(59) \cdot 10^{-3} ~ , ~ \\[2mm]
%     \label{eq:chi1-_final}
%     \chi_{1^-}(m_b^{phys}) & = & 5.84\,(44) \cdot 10^{-4}~{\rm GeV}^{-2} ~ , ~ \\[2mm]
%     \label{eq:chi0-_final}
%     \chi_{0^-}(m_b^{phys}) & = & 2.19\,(19) \cdot 10^{-2} ~ , ~ \\[2mm]
%     \label{eq:chi1+_final}
%     \chi_{1^+}(m_b^{phys}) & = & 4.69\,(30) \cdot 10^{-4}~{\rm GeV}^{-2} ~ , ~
% \eea
which represent our non-perturbative determinations of the dispersive bounds on the form factors of the exclusive semileptonic $B \to D^{(*)} \ell \nu_\ell$ decays. 

The application of the above results to the extraction of the CKM matrix element $|V_{cb}|$ from the exclusive experimental data will be illustrated in a separate work~\cite{Martinelli:2021onb}.

Finally, we mention that in this work we have limited ourselves to evaluate the susceptibilities at zero momentum transfer.
However, our non-perturbative calculations of the relevant two-point correlation functions allow to consider the most convenient value of the 4-momentum transfer leading to the most stringent bounds on the semileptonic form factors. We leave this investigation to a future work.

\section*{Acknowledgments}
We acknowledge PRACE for awarding us access to Marconi at CINECA, Italy under the grant the PRACE projects PRA027 and PRA067.  We also acknowledge use of CPU time provided by CINECA under the specific initiative INFN-LQCD123.
G.M. and S.S. thank MIUR (Italy) for partial  support under the contract PRIN 2015. 
S.S is supported by the Italian Ministry of Research (MIUR) under grant PRIN 20172LNEEZ.

\appendix

\section{Simulation details}
\label{sec:simulations}

The ETMC setup adopted in this work is based on the Iwasaki action \cite{Iwasaki:1985we} for the gluons and on the Wilson maximally twisted-mass action \cite{Frezzotti:2000nk,Frezzotti:2003xj,Frezzotti:2003ni} for the sea quarks. 
Three values of the inverse bare lattice coupling $\beta$ and different lattice volumes are considered, as it is shown in Table \ref{tab:simudetails}, where the number of configurations analyzed ($N_{cfg}$) corresponds to a separation of $20$ trajectories.

At each lattice spacing different values of the light sea quark mass are considered, and the light valence and sea quark masses are always taken to be degenerate, i.e.~$m_{ud}^{sea} = m_{ud}^{val} = m_{ud}$. 
In order to avoid the mixing of strange and charm quarks in the valence sector we adopt a non-unitary set up in which the valence strange and charm quarks are regularized as Osterwalder-Seiler fermions \cite{Osterwalder:1977pc}, while the valence up and down quarks have the same action of the sea.
Working at maximal twist such a setup guarantees an automatic ${\cal{O}}(a)$-improvement \cite{Frezzotti:2003ni,Frezzotti:2004wz}.
Quark masses are renormalized through the RC $Z_m = 1 / Z_P$, computed non-perturbatively using the RI$^\prime$-MOM scheme (see Ref.~\cite{Carrasco:2014cwa}).

The lattice scale is determined using the experimental value of $f_{\pi^+}$ so that the values of the lattice spacing are $a = 0.0885(36), ~ 0.0815(30),  ~ 0.0619(18)$ fm at $\beta = 1.90, ~ 1.95$ and $2.10$, respectively, the lattice size goes from $\simeq 2$ to $\simeq 3$ fm.

The physical up/down, strange and charm quark masses are obtained by using the experimental values for $M_\pi$, $M_K$ and $M_{D_s}$, obtaining~\cite{Carrasco:2014cwa} $m_{ud}^{phys} = 3.72 \pm 0.17$ MeV, $m_s^{phys} = 99.6 \pm 4.3$ MeV and $m_c^{phys} = 1.176 \pm 0.039$ GeV in the $\overline{\mathrm{MS}}$ scheme at a renormalization scale of 2 GeV.
%The values of the strange and charm sea-quark masses corresponding to the ETMC ensembles of Table \ref{tab:simudetails} have been calculated in Ref.~\cite{Carrasco:2014cwa}, obtaining a slight mistuning.
%It was shown~\cite{Carrasco:2014cwa} that such a mistuning may produce changes in the determination of the physical quark masses smaller than other systematic uncertainties.
In Ref.~\cite{Bussone:2016iua} the physical b-quark mass is determined adopting the ETMC ratio method~\cite{Blossier:2009hg}, obtaining $m_b^{phys}(m_b^{phys}) = 4.26 \pm 0.10$ GeV which corresponds to $m_b^{phys} = 5.198 \pm 0.122$ GeV in the $\overline{\mathrm{MS}}(2~\mbox{GeV})$ scheme.

We have considered three values of the valence-quark bare mass in both the charm and the strange sectors, which are needed to interpolate smoothly to the corresponding physical strange and charm regions.
For each lattice spacing the bare masses are chosen so that the corresponding renormalized masses are in the following ranges: $3 m_{ud}^{phys} \lesssim m_{ud} \lesssim 12 m_{ud}^{phys}$, $0.7 m_s^{phys} \lesssim m_s \lesssim 1.2 m_s^{phys}$ and $0.7 m_c^{phys} \lesssim m_c \lesssim 1.1 m_c^{phys}$. 
In order to extrapolate up to the $b$-quark sector we have also considered seven values of the valence heavy-quark mass, $m_h$, in the range $1.1 m_c^{phys} \lesssim m_h \lesssim 3.3 m_c^{phys} \approx 0.8 m_b^{phys}$. 

\begin{table}[htb!]
{\small 
\begin{center}
\begin{tabular}{||c|c|c|c||c|c|c|c||c||}
\hline
ensemble & $\beta$ & $V / a^4$ &$N_{cfg}$&$a\mu_{ud}$& $a\mu_s$ & $a\mu_c$ & $a\mu_h > a \mu_c$ & $M_\pi$ (MeV) \\
\hline \hline
$A30.32$ & $1.90$ & $32^3\times 64$ & $150$ & $0.0030$ & $\{0.0180,$ & $\{0.21256,$ & $\{0.34583, 0.40675,$ & 275~(10)\\
$A40.32$ & & & $150$ & $0.0040$ & $0.0220,$ & $~0.25000,$ & $~0.47840, 0.56267,$ & 316~(12) \\
$A50.32$ & & & $150$ & $0.0050$ & $0.0260\}$ & $~~0.29404\}$ & $~0.66178, 0.77836,$ & 350~(13) \\
\cline{1-1} \cline{3-5}
$A40.24$ & & $24^3\times 48 $ & $150$ & $0.0040$ & & & $~0.91546\},$ & 322~(13) \\
$A60.24$ & & & $150$ & $0.0060$ & & & & 386~(15)\\
$A80.24$ & & & $150$ & $0.0080$ & & & & 442~(17) \\
$A100.24$ & & & $150$ & $0.0100$ & & & & 495~(19) \\
\hline \hline
$B25.32$ & $1.95$ & $32^3\times 64$ & $150$ & $0.0025$ & $\{0.0155,$ & $\{0.18705,$ & $\{0.30433, 0.35794,$ & 259~~(9) \\
$B35.32$ & & & $150$ & $0.0035$ & $0.0190,$ & $~0.22000,$ & $~0.42099, 0.49515,$ & 302~(10) \\
$B55.32$ & & & $150$ & $0.0055$ & $0.0225\}$ & $~~0.25875\}$ & $~0.58237, 0.68495,$ & 375~(13) \\
$B75.32$ & & & $~75$ & $0.0075$ & & & $~0.80561\}$ & 436~(15) \\
\cline{1-1} \cline{3-5}
$B85.24$ & & $24^{3}\times 48 $ & $150$ & $0.0085$ & & & &468~(16) \\
\hline \hline
$D15.48$ & $2.10$ & $48^3\times 96$ & $~90$ & $0.0015$ & $\{0.0123,$ & $\{0.14454,$ & $\{0.23517, 0.27659,$ & 223~(6) \\ 
$D20.48$ & & & $~90$& $0.0020$ & $0.0150,$ & $~0.17000,$ & $~0.32531, 0.38262,$ & 256~(7) \\
$D30.48$ & & & $~90$& $0.0030$ & $0.0177\}$ & $~~0.19995\}$ & $~0.45001, 0.52928,$ & 312~(8) \\
                 & & &           &                 &                   &                         & $~0.62252\}$ & \\
\hline   
\end{tabular}
\end{center}
}
\vspace{-0.250cm}
\caption{\it \small Values of the valence-quark bare masses in the light ($a \mu_{ud}$), strange ($a \mu_s$), charm ($a \mu_c$) and heavier-than-charm ($a \mu_h$) regions considered for the $15$ ETMC gauge ensembles with $N_f = 2+1+1$ dynamical quarks (see Ref.~\cite{Carrasco:2014cwa}). $N_{cfg}$ stands for the number of (uncorrelated) gauge configurations used in this work. In the last column the values of the simulated pion mass, $M_\pi$, are shown in physical units.}
\label{tab:simudetails}
\end{table}

The statistical accuracy of the meson correlators (\ref{eq:CVL12}-\ref{eq:CP12}) can be significantly improved by adopting the ``one-end" trick stochastic method~\cite{Foster:1998vw, McNeile:2006bz}, which employs spatial stochastic sources at a single time slice chosen randomly.

In Ref.~\cite{Carrasco:2014cwa} the analysis is split into eight branches, which differ in: 
\begin{itemize}
\item the continuum extrapolation adopting for the matching of the lattice scale either the Sommer parameter $r_0$ or the mass of a fictitious P-meson made up of two valence strange(charm)-like quarks; 
\item the chiral extrapolation performed with fitting functions chosen to be either a polynomial expansion or a Chiral Perturbation Theory (ChPT) Ansatz in the light-quark mass;
\item the choice between the methods M1 and M2, which differ by ${\cal{O}}(a^2)$ effects, used to determine the mass RC $Z_m = 1 / Z_P$ in the RI$^\prime$-MOM scheme. 
\end{itemize}
In the present analysis we will make use of the input parameters corresponding to each of the eight branches of Ref.~\cite{Carrasco:2014cwa}.
For each branch the central values and the errors of the input parameters are evaluated using a bootstrap sample with ${\cal{O}}(100)$ events. The corresponding results are collected in Tables~\ref{tab:8branches} and \ref{tab:RCs}.

\begin{table}[htb!]
{\footnotesize
\begin{center}
\begin{tabular}{||c|l ||c|c|c|c||c||} 
\hline 
\multicolumn{1}{||c}{} & \multicolumn{1}{|c||}{$\beta$} & \multicolumn{1}{c|}{ $1^{st}$ } & \multicolumn{1}{c|}{ $2^{nd}$ } & \multicolumn{1}{c|}{ $3^{rd}$ } & \multicolumn{1}{c||}{ $4^{th}$ } & \multicolumn{1}{c||}{ $1^{st} - 4^{th}$ } \\ \hline \hline   
                                  & 1.90 & 2.224(68) & 2.192(75) & 2.269(86) & 2.209(84) & 2.224(84) \\ 
$a^{-1}({\rm GeV})$  & 1.95 & 2.416(63) & 2.381(73) & 2.464(85) & 2.400(83) & 2.415(82) \\ 
                                 & 2.10 & 3.184(59) & 3.137(64) & 3.248(75) & 3.163(75) & 3.183(80) \\ \hline \hline
$m_{ud}^{phys}({\rm GeV})$ & & 0.00372(13) & 0.00386(17) & 0.00365(10) & 0.00375(13) & 0.00375(16) \\ \hline
$m_s^{phys}$({\rm GeV})      & & 0.1014(43)   & 0.1023(39)   & 0.0992(29)   & 0.1007(32)   & 0.1009(38) \\ \hline
$m_c^{phys}$({\rm GeV})      & & 1.183(34)     & 1.193(28)     & 1.177(25)     & 1.219(21)     & 1.193(32) \\ \hline \hline
\end{tabular}
\\ \vspace{0.25cm}
\begin{tabular}{||c|l ||c|c|c|c||c||} 
\hline 
\multicolumn{1}{||c}{} & \multicolumn{1}{|c||}{$\beta$} & \multicolumn{1}{c|}{ $5^{th}$ } & \multicolumn{1}{c|}{ $6^{th}$ } & \multicolumn{1}{c|}{ $7^{th}$ } & \multicolumn{1}{c||}{ $8^{th}$ } & \multicolumn{1}{c||}{ $5^{th} - 8^{th}$ } \\ \hline \hline  
                                  & 1.90 & 2.222(67) & 2.195(75) & 2.279(89) & 2.219(87) & 2.229(86) \\ 
$a^{-1}({\rm GeV})$  & 1.95 & 2.414(61) & 2.384(73) & 2.475(88) & 2.411(86) & 2.421(85) \\ 
                                 & 2.10 & 3.181(57) & 3.142(64) & 3.262(79) & 3.177(78) & 3.191(83) \\ \hline \hline
$m_{ud}^{phys}({\rm GeV})$ & & 0.00362(12) & 0.00377(16) & 0.00354(9) & 0.00363(12) & 0.00364(15) \\ \hline
$m_s^{phys}({\rm GeV})$      & & 0.0989(44)   & 0.0995(39)   & 0.0962(27) & 0.0975(30)  & 0.0980(38) \\ \hline
$m_c^{phys}({\rm GeV})$      & & 1.150(35)     & 1.158(27)     & 1.144(29)   & 1.182(19)    & 1.159(32) \\ \hline \hline   
\end{tabular} 
\end{center}
}
\vspace{-0.25cm}
\caption{\it \small The input parameters for the eight branches of the analysis of Ref.~\cite{Carrasco:2014cwa}. The renormalized quark masses are given in the $\overline{\mathrm{MS}}$ scheme at a renormalization scale of 2 GeV. The last columns represent the averages of the previous four columns (according to Eq.~(28) of Ref.~\cite{Carrasco:2014cwa}). With respect to Ref.~\cite{Carrasco:2014cwa} the table includes an update of the values of the lattice spacing and, consequently, of all the other quantities.}
\label{tab:8branches}
\end{table} 

\begin{table}[htb!]
{\small
\begin{center}
\begin{tabular}{||c|lc|c|c||c|c|c||} 
\hline 
\multicolumn{1}{||c}{} & \multicolumn{3}{||c||}{ $1^{st} - 4^{th}$ branches} & \multicolumn{3}{|c||}{ $5^{th} - 8^{th}$ branches} \\ \cline{2-7}
\multicolumn{1}{||c}{} & \multicolumn{1}{||c|}{ $\beta = 1.90$ } & \multicolumn{1}{c|}{ $\beta = 1.95$ } & \multicolumn{1}{c||}{ $\beta = 2.10$ } 
                                  & \multicolumn{1}{c|}{ $\beta = 1.90$ } & \multicolumn{1}{c|}{ $\beta = 1.95$ } & \multicolumn{1}{c||}{ $\beta = 2.10$ } \\ \hline \hline
\multicolumn{1}{||c}{$Z_V$} & \multicolumn{1}{||c|}{0.5920(4)} & \multicolumn{1}{|c|}{0.6095(3)} & \multicolumn{1}{|c||}{0.6531(2)}
                                             & \multicolumn{1}{|c|}{0.5920(4)} & \multicolumn{1}{|c|}{0.6095(3)} & \multicolumn{1}{c||}{0.6531(2)} \\ \hline
\multicolumn{1}{||c}{$Z_A$} & \multicolumn{1}{||c|}{0.731(8)} & \multicolumn{1}{|c|}{0.737(5)} & \multicolumn{1}{|c||}{0.762(4)}
                                             & \multicolumn{1}{|c|}{0.703(2)} & \multicolumn{1}{|c|}{0.714(2)} & \multicolumn{1}{c||}{0.752(2)} \\ \hline
\multicolumn{1}{||c}{$Z_P$} & \multicolumn{1}{||c|}{0.529(7)} & \multicolumn{1}{|c|}{0.509(3)} & \multicolumn{1}{|c||}{0.516(3)}
                                             & \multicolumn{1}{|c|}{0.573(4)} & \multicolumn{1}{|c|}{0.544(2)} & \multicolumn{1}{c||}{0.542(1)} \\ \hline
\multicolumn{1}{||c}{$Z_S$} & \multicolumn{1}{||c|}{0.747(12)} & \multicolumn{1}{|c|}{0.713(9)} & \multicolumn{1}{|c||}{0.700(6)}
                                             & \multicolumn{1}{|c|}{0.877(3)} & \multicolumn{1}{|c|}{0.822(2)} & \multicolumn{1}{c||}{0.749(3)} \\ \hline
\end{tabular}
\end{center}
}
\vspace{-0.25cm}
\caption{\it \small Values of the RCs for the eight branches of the analysis evaluated in Ref.~\cite{Carrasco:2014cwa} using the vector Ward-Takahashi identity for $Z_V$ and the RI$^\prime$-MOM scheme for all the others. The scale-dependent RCs $Z_P$ and $Z_S$ are given in the $\overline{\mathrm{MS}}$ scheme at a renormalization scale of 2 GeV.}
\label{tab:RCs}
\end{table} 

Besides the RC $Z_P$ we need the RCs of other bilinear quark operators, namely $Z_V$, $Z_A$ and $Z_S$ related respectively to the vector, axial-vector and scalar currents. 
They have been evaluated in the Appendix of Ref.~\cite{Carrasco:2014cwa} in the RI$^\prime$-MOM scheme for $Z_A$ and $Z_S$, while for $Z_V$ we adopt its determination based on the vector Ward-Takahashi identity.

Throughout this work\footnote{Unless otherwise stated, the results that will be shown in all the Figures correspond to the average of the first four branches of the bootstrap analysis.} the results ($x_i \pm \sigma_i$ ) with $i = 1, 2, ... N$, obtained within $N$ branches, are averaged according to the following general formula (see Ref.~\cite{Carrasco:2014cwa})
\bea
    \overline{x} & = & \frac{1}{N} \sum_{i=1}^N x_i ~ , \nonumber \\ 
    \sigma^2 & = & \frac{1}{N} \sum_{i=1}^N \sigma_i^2  + \frac{1}{N} \sum_{i=1}^N (x_i  - \overline{x})^2 ~ .
    \label{eq:combineresults}
 \eea
The second term in the r.h.s.~of Eq.~(\ref{eq:combineresults}), coming from the spread among the results of the different branches, corresponds to the systematic error which accounts for the uncertainties due to the chiral extrapolation, the cutoff effects and the choice of the RC $Z_P$.

\end{document}